\documentclass{article}

\usepackage{arxiv}

\usepackage{amsmath}
\usepackage{amssymb}
\usepackage{amsfonts}
\usepackage{latexsym}
\usepackage{bbm}
\usepackage{bm}

\usepackage{makecell}
\usepackage{booktabs}
\usepackage{caption} 
\usepackage{multirow}

\usepackage[utf8]{inputenc} 
\usepackage[T1]{fontenc}    
\usepackage{hyperref}       
\usepackage{url}            
\usepackage{nicefrac}       
\usepackage{microtype}      
\usepackage{lipsum}		    
\usepackage{graphicx}
\usepackage{doi}
\usepackage{xcolor}
\usepackage{csquotes}

\graphicspath{{./figures/}}

\title{CartiMorph: a framework for automated knee articular cartilage morphometrics\thanks{This preprint is an proofread version of a paper published in Medical Image Analysis, which can be found at https://doi.org/10.1016/j.media.2023.103035}}

\date{} 					

\author{ 
\href{https://orcid.org/0000-0003-2754-3649}{\includegraphics[scale=0.06]{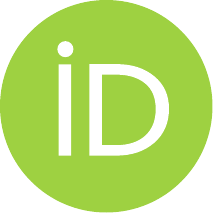}}\hspace{1mm}Yongcheng Yao$^{1\S}$, 
\href{https://orcid.org/0000-0002-3897-9280}{\includegraphics[scale=0.06]{orcid.pdf}}\hspace{1mm}Junru Zhong$^{1\dag}$, 
\href{https://orcid.org/0000-0003-1962-0106}{\includegraphics[scale=0.06]{orcid.pdf}}\hspace{1mm}Liping Zhang$^1$\thanks{equal contribution}, \href{https://orcid.org/0000-0002-1975-4334}{\includegraphics[scale=0.06]{orcid.pdf}}\hspace{1mm}Sheheryar Khan$^2$, 
\href{https://orcid.org/0000-0001-7242-9285}{\includegraphics[scale=0.06]{orcid.pdf}}\hspace{1mm}Weitian Chen$^1$\thanks{corresponding author} \\ \\
$^1$ CU Lab of AI in Radiology (CLAIR),\\ Department of Imaging and Interventional Radiology, \\The Chinese University of Hong Kong \\ \\
$^2$ School of Professional Education and Executive Development,\\ The Hong Kong Polytechnic University \\ \\
\texttt{$^{\S}$yao\_yongcheng@link.cuhk.edu.hk; $^{\ddag}$wtchen@cuhk.edu.hk}
}

\date{}

\begin{document}
\maketitle
\begin{abstract}
    We introduce CartiMorph, a framework for automated knee articular cartilage morphometrics. It takes an image as input and generates quantitative metrics for cartilage subregions, including the percentage of full-thickness cartilage loss (FCL), mean thickness, surface area, and volume. CartiMorph leverages the power of deep learning models for hierarchical image feature representation. Deep learning models were trained and validated for tissue segmentation, template construction, and template-to-image registration. We established methods for surface-normal-based cartilage thickness mapping, FCL estimation, and rule-based cartilage parcellation. Our cartilage thickness map showed less error in thin and peripheral regions. We evaluated the effectiveness of the adopted segmentation model by comparing the quantitative metrics obtained from model segmentation and those from manual segmentation. The root-mean-squared deviation of the FCL measurements was less than 8\%, and strong correlations were observed for the mean thickness (Pearson's correlation coefficient $\rho \in [0.82,0.97]$), surface area ($\rho \in [0.82,0.98]$) and volume ($\rho \in [0.89,0.98]$) measurements. We compared our FCL measurements with those from a previous study and found that our measurements deviated less from the ground truths. We observed superior performance of the proposed rule-based cartilage parcellation method compared with the atlas-based approach. CartiMorph has the potential to promote imaging biomarkers discovery for knee osteoarthritis.
\end{abstract}


\section{Introduction}
\label{sec:intro}
Knee articular cartilage morphometrics is an effective tool to derive imaging
biomarkers for knee osteoarthritis
(OA)\cite{gray2004toward,mosher2011knee,hunter2014biomarkers,eckstein2015brief,collins2016semiquantitative,guermazi2017brief,wirth2017predictive,everhart2019full,hunter2022multivariable}.
Magnetic resonance (MR) imaging is a predominant non-invasive imaging modality
for OA research\cite{oei2021osteoarthritis} because of its ability to
anatomically and biochemically quantify alterations in the cartilage at
different stages of OA progression. From a structural MR image, morphological
biomarkers can be analyzed and correlated with clinical outcomes. Tremendous
efforts have been made to develop automated analysis techniques for imaging
biomarker extraction. On the one hand, semiautomated methods for the
quantification of cartilage lesions, thickness, surface area, and volume have
been
proposed\cite{cohen1999knee,stammberger1999determination,hohe2002surface,kauffmann2003computer,eckstein2006magnetic,eckstein2006double,carballido2008inter,wirth2008technique,williams2010anatomically,maerz2016surface,favre2017anatomically}.
On the other hand, semiquantitative methods for the grading of cartilage
lesions have also been
proposed\cite{eckstein2006magnetic,guermazi2017brief,dorio2020association}.
These semiquantitative methods usually rely on MR-based grading systems, such
as WORMS\cite{peterfy2004whole}, KOSS\cite{kornaat2005mri},
BLOKS\cite{hunter2008reliability}, MOAKS\cite{hunter2011evolution}, and
CaLS\cite{alizai2014cartilage}. Semiautomated and semiquantitative methods
are subjective, error-prone, and time-consuming as they require human
interaction and domain knowledge. To overcome such drawbacks,  automated tissue
segmentation methods have been extensively exploited and
validated\cite{prasoon2013deep,raj2018automatic,liu2018deep,zhou2018deep, ambellan2019automated,tan2019collaborative,xu2019deepatlas,gaj2020automated,khan2022deep}.
To the best of our knowledge, there is a lack of automated methods for the
quantification of cartilage lesions. To fill this gap, we aimed to automate the
quantitative measurement of full-thickness cartilage loss (FCL) which has been
shown to be an imaging biomarker for OA
progression\cite{guermazi2017brief,everhart2019full}. Additionally, we
developed methods for robust cartilage thickness mapping, rule-based cartilage
parcellation, and regional quantification. 

In this work, we developed a framework called CartiMorph for automated knee articular cartilage morphometrics. The quantitative metrics for cartilage subregions include the percentage of FCL, mean thickness, surface area, and volume. Deep learning models were integrated into the proposed framework for tissue segmentation, template construction, and template-to-image registration. Our code and experimental details are publicly available.\footnote[1]{Available at https://github.com/YongchengYAO/CartiMorph.} The contributions of this work are summarized as follows.

\begin{enumerate}
  \item We established a method for automated cartilage thickness mapping that is robust to cartilage lesions.
  \item We established a method for automated FCL estimation through learning-based deformable image registration, template construction, and surface-based operations.
  \item We established a rule-based cartilage parcellation method that automatically partitions femoral and tibial cartilages into 20 subregions. The parcellation method was designed for robust regional quantification despite the extent of the FCL.
  \item We constructed a knee template image learned from MR scans of subjects without radiographic OA. We additionally constructed the respective 5-class tissue segmentation and a 20-region atlas for the template image, which provide prior knowledge of the bone and cartilage anatomy.
  \item We obtained new insights into the effectiveness of deep-learning-based segmentation models in cartilage morphometrics.
\end{enumerate}

The remainder of this article is organized as follows.  \ref{sec:relatedWorks} describes related works on tissue segmentation \ref{subsec:relatedWorks-img-seg}, image registration \ref{subsec:relatedWorks-img-reg}, template construction \ref{subsec:relatedWorks-template}, and cartilage morphometrics \ref{subsec:relatedWorks-CartiMorph}.  \ref{sec:method} provides details of the proposed framework.  \ref{sec:experiment} presents the experiments and results.  \ref{sec:discussion} discusses the insights gained from the results, the limitations of this work, and future works. We draw conclusions from the results in  \ref{sec:conclusion}.

\section{Related Works}
\label{sec:relatedWorks}

\subsection{Tissue Segmentation}
\label{subsec:relatedWorks-img-seg}
Tissue segmentation is a vital upstream task in imaging biomarker analysis.
Some downstream tasks, such as morphological quantification, rely heavily on
the accuracy of tissue segmentation. In knee OA, the articular
cartilage\cite{eckstein2015brief,guermazi2017brief,wirth2017predictive,everhart2019full}
and
bone\cite{bredbenner2010statistical,driban2013evaluation,neogi2013magnetic,barr2015systematic,bowes2015novel,hunter2016longitudinal,barr2016relationship,mackay2016mri,morales2020learning}
are two regions of interest (ROI) where morphological alterations relate to
disease progression. Segmentation methods can be categorized into
non-machine-learning, classical machine learning, and deep learning approaches.
Non-machine-learning segmentation techniques include deformable-model-based,
graph-based, and atlas-based methods. These methods either rely on
heuristically designed algorithms or require user interaction. Classical
machine learning methods include decision tree, Bayes classifier, support
vector machine, clustering, ensemble learning, and reinforcement learning.
These methods often need extensive feature engineering which involves feature
extraction and selection. Deep learning is a subfield of machine learning
wherein models implicitly learn features from data. Deep learning has attracted
strong interest for its superior performance and fast inference property.

Recent advances in deep-learning-based knee articular cartilage and bone
segmentation\cite{gan2021classical} are summarized as follows. Convolutional
neural networks (CNN) are widely used in semantic segmentation. Triplanar
two-dimensional (2D) CNN\cite{prasoon2013deep} was applied to cartilage
segmentation in an early study. U-Net\cite{ronneberger2015u,cciccek20163d}, a
CNN with an encoder--decoder architecture and shortcut connections, was
proposed for biomedical image segmentation.
$\mu$-Net\cite{raj2018automatic}, an early work on cartilage
segmentation using three-dimensional (3D) CNN, featured short residual
connections\cite{he2016deep}, long skip connections\cite{ronneberger2015u},
deep supervision\cite{lee2015deeply}, and extra input feeding. The
feasibility of refining the segmentation output via the conditional random
field, simplex deformable model, and statistical shape model has been
explored\cite{liu2018deep,zhou2018deep,ambellan2019automated}. Although shape
models can improve segmentation accuracy, they are difficult to integrate into
an end-to-end training framework. To this end, complicated network structures
and training strategies have been
developed\cite{tan2019collaborative,xu2019deepatlas,gaj2020automated}. A
recent work demonstrated the feasibility of combining CNN, block-wise low-rank
reconstruction, and alpha matting for cartilage
segmentation\cite{khan2022deep}. However, the use of complex networks may
face difficulty in training and poor generalizability. Methods evaluated on one
dataset may fail to transfer to another dataset owing to the diversity of data
characteristics (\emph{i.e.}, the domain shift problem). It would be desirable
to have an efficient self-configuring segmentation framework that works on a
wide variety of datasets. nnU-Net\cite{isensee2021nnu}, an
auto-machine-learning\cite{hutter2019automated} framework, has performed well
in medical image segmentation, outperforming most state-of-the-art methods on
23 public datasets and 53 segmentation tasks that cover multiple target
structures and image modalities.

\subsection{Image Registration}
\label{subsec:relatedWorks-img-reg}
Image registration is a process of warping a moving image to align with a fixed image so that the two images have similar appearances at the same locations. Essentially, image registration is an optimization of image transformation. Mathematically, the optimization process can be formulated as the minimization of an objective function that quantifies image dissimilarity and regularizes the deformation field. Transformations can be classified as affine and deformable transformations in terms of their complexity. As the affine transformations are too simplistic for most real-world registration problems, the deformable methods are more extensively exploited. Deformable registrations can be categorized into non-learning-based and learning-based methods. Non-learning-based registration continues iterative optimization for each pair of images until convergence, including Demons \cite{thirion1998image, vercauteren2009diffeomorphic}, HAMMER \cite{shen2007image}, SyN \cite{avants2008symmetric}, and Elastix \cite{klein2009elastix}. In contrast, learning-based registration methods learn a global parameterized function from data and use the estimated function for fast image registration.  Recently, supervised and unsupervised deep learning methods have shown comparable or better performance than conventional methods while being orders of magnitude faster \cite{fu2020deep, zou2022review}.

Recent advances in deep learning methods for image registration are summarized as follows. Deformable registration via supervised learning requires ground truth deformation fields which can be 1) random fields, 2) model-generated fields, or 3) traditional-registration-generated fields. Randomly synthesized deformation vector field (DVF) (\emph{i.e.}, dense displacement field) has been used for supervised registration \cite{sokooti2017nonrigid, eppenhof2018pulmonary}. Since the randomly synthesized DVF may fail to mimic true deformations, model-generated \cite{uzunova2017training, sokooti20193d} and traditional-registration-generated \cite{cao2017deformable, fan2019birnet} artificial DVFs have been used. In the former case, synthetic DVFs and respective image pairs are generated using certain deformation models. In the latter case, conventional non-learning-based registration methods are used to estimate DVFs from real image pairs. The spatial transformer network (STN) \cite{jaderberg2015spatial} helped catalyze the development of  unsupervised deformable registration. STN introduced a differentiable spatial transformer that learns transformations for feature maps to improve the spatial invariance of models. Inspired by STN, the spatial transformer (\emph{i.e.}, spatial transformation layer) has been adopted extensively in unsupervised registration works \cite{vos2017end, yoo2017ssemnet, li2018non, de2019deep}. VoxelMorph \cite{balakrishnan2019voxelmorph} was evaluated on a large multi-study brain MR dataset and found to have performance comparable to that of SyN (implemented in ANTs \cite{avants2011reproducible}) and NiftyReg \cite{modat2010fast}. VoxelMorph was further extended to a probabilistic framework \cite{dalca2019unsupervised}. Other techniques for boosting the registration accuracy include the joint learning of segmentation and registration tasks \cite{qin2018joint, xu2019deepatlas}, the diffeomorphic registration \cite{dalca2019unsupervised, krebs2019learning, mok2020fast}, the inverse-consistency constraint \cite{zhang2018inverse, mok2020fast}, the cycle-consistency constraint \cite{kim2021cyclemorph}, and transformer-based architecture \cite{chen2022transmorph, shi2022xmorpher, zhu2022swin}. Although deep-learning-based image registration is a rapidly developing field, few methods of such registration have been evaluated on knee MR images \cite{xu2019deepatlas, shen2019networks, ding2022aladdin}.

\subsection{Template Construction}
\label{subsec:relatedWorks-template}
In the literature, the terms \enquote{template} and \enquote{atlas} are used
interchangeably, and they can have different meanings across fields. To avoid
ambiguity, we herein define the template as an \enquote{average} image and the
atlas as a subregion parcellation built on the template image. The template
image is useful for morphological analyses as it defines a reference space for
groupwise comparison and provides prior information on structures of interest.
In general, template construction refers to building a representative image by
aligning a set of images, and it thus naturally relates to image registration.
Similar to registration, template construction can be categorized into
non-learning-based and learning-based methods. Conventional
methods\cite{joshi2004unbiased,ashburner2007fast} involve an iterative
optimization-based registration process whereby each image is aligned with a
tentative template that is updated at the end of each iteration. Such an
approach is prohibitively expensive in computation. With the success of deep
learning in computer vision, there has been extensive methodological
development in image registration and synthesis, nurturing learning-based
template construction techniques. A learning-based template building
method\cite{dalca2019learning} was proposed for simultaneous diffeomorphic
registration and conditional template generation. More recent improvements
include the use of generative adversarial
network\cite{goodfellow2014generative,dey2021generative}, the joint learning
of segmentation and registration\cite{he2021learning,sinclair2022atlas}, the
temporal-consistency constraint\cite{chen2021construction}, and the
anatomical constraint\cite{pei2021learning}.

\subsection{Cartilage Morphometrics}
\label{subsec:relatedWorks-CartiMorph}

\subsubsection{Cartilage Thickness Mapping}
\label{subsubsec:relatedWorks-thickness}
Cartilage thickness mapping refers to measuring the thickness across the entire
cartilage surface. Despite the variation in measuring methods, defining a
surface onto which thickness values are mapped is always the first step. Since
the bone--cartilage interface (\emph{i.e.}, inner surface) is subject to less
shape variation across time than the outer surface, it has been chosen as the
target surface for thickness mapping in most
studies\cite{kauffmann2003computer,carballido2008inter,williams2010anatomically,favre2017anatomically}.
Early \emph{in vivo} thickness mapping
works\cite{cohen1999knee,kauffmann2003computer,carballido2008inter}
delineated the inner and outer boundaries of cartilage from MR images using
semiautomated edge detection methods such as active contour. Later
works\cite{wirth2008technique,williams2010anatomically,favre2013patterns,maerz2016surface,van2017knee}
leveraged the rich spatial information in manual pixel-wise segmentation for
surface-based analyses. With the emergence of efficient segmentation models,
recent
studies\cite{norman2018use,si2021knee,wirth2021accuracy,panfilov2022deep,eckstein2022detection}
have evaluated the effectiveness of such automated segmentation methods in
cartilage thickness mapping.

The implementations of thickness measurement methods vary across studies and we
only discuss the mathematical definitions here. Two prevalent methods are the
nearest neighbor
approach\cite{carballido2008inter,favre2013patterns,favre2017anatomically}
which calculates thickness as the minimal Euclidean distance between two
surfaces, and the surface normal
approach\cite{williams2010anatomically,maerz2016surface,van2017knee} which
measures the Euclidean distance in the normal direction of the target surface.
Other methods are based on the local thickness\cite{hildebrand1997new}, an
electrostatic model (\emph{i.e.}, Laplace's equation
method)\cite{jones2000three}, and the Eulerian partial differential
equation\cite{yezzi2003eulerian}.

\subsubsection{Full-thickness Cartilage Loss Estimation}
\label{subsubsec:relatedWorks-FCL}
Semiautomated methods\cite{eckstein2006double,wirth2008technique} have been
proposed for the regional quantification of cartilage lesions, whereas
semiquantitative methods\cite{guermazi2017brief,dorio2020association} have
been adopted for the severity grading of cartilage lesions. There are few
automated methods for FCL quantification.

\subsubsection{Cartilage Parcellation}
\label{subsubsec:relatedWorks-parcellation}
Partitioning cartilage into subregions enables regional analysis of
quantitative metrics. A variety of cartilage parcellation schemes have been
proposed in previous
studies\cite{kauffmann2003computer,koo2005considerations,pelletier2007risk,wirth2008technique,wirth2010spatial,surowiec2014t2}.
Additionally, the semiquantitative scoring system
WORMS\cite{peterfy2004whole} provides a division scheme. Cartilage
parcellation algorithms can be classified into
manual\cite{kauffmann2003computer},
rule-based\cite{koo2005considerations,wirth2008technique,wirth2010spatial,surowiec2014t2},
and atlas-based methods\cite{panfilov2022deep}. Rule-based methods rely on
parcellation rules, whereas atlas-based methods warp a predefined atlas to
individual segmentation via image registration. Rule-based methods are
preferred when the parcellation rules can be turned into robust algorithms, and
atlas-based methods are more feasible otherwise. Atlas-based methods can
achieve satisfactory performance when the subregions are defined by
distinguishable anatomical landmarks because such landmarks can implicitly
improve image registration. However, there is a lack of precise anatomical
landmarks and boundaries between subregions defined in most cartilage
parcellation schemes\cite[Fig. 1]{favre2017anatomically}. Owing to the simple
anatomy of the knee articular cartilage and the lack of distinguishable
landmarks for all subregions, rule-based methods are more suitable in cartilage
morphometrics.

\section{Method}
\label{sec:method}

\begin{figure}[!t]
\centerline{\includegraphics[width=1\columnwidth]{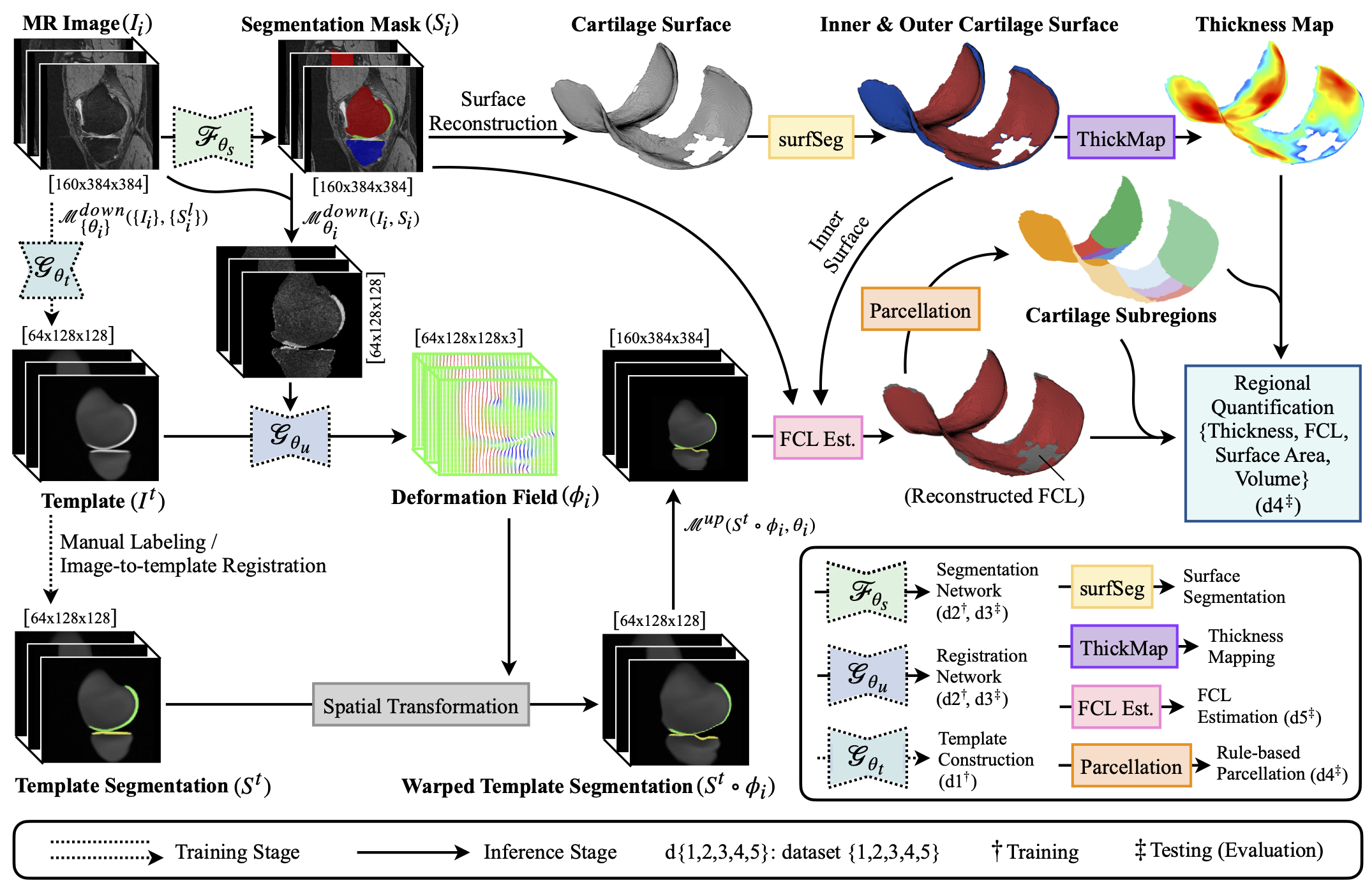}}
\caption{Schematic overview of the proposed framework. A knee template image $\boldsymbol{I}^t$ is learned in the network $\mathcal{G}_{\boldsymbol{\theta}_t}$ from low-resolution images masked by segmentation labels $\{\boldsymbol{I}^{\text{low}}_i\}=\mathcal{M}^\text{down}_{\{\boldsymbol{\theta}_i\}}(\{\boldsymbol{I}_i\},\{\boldsymbol{S}^l_i\})$. The template segmentation $\boldsymbol{S}^t$ is constructed via manual labeling or image-to-template registration. Both the template image $\boldsymbol{I}^t$ and the template segmentation $\boldsymbol{S}^t$ are constructed once in the training stage and used for every input image in the inference stage. The segmentation network $\mathcal{F}_{\boldsymbol{\theta}_s}$ predicts the bone and cartilage segmentation masks $\boldsymbol{S}_i$ for the image $\boldsymbol{I}_i$. The template image $\boldsymbol{I}^t$ is registered to the low-resolution image $\boldsymbol{I}^{\text{low}}_i=\mathcal{M}^\text{down}_{\boldsymbol{\theta}_i}(\boldsymbol{I}_i,\boldsymbol{S}_i)$ via a registration network $\mathcal{G}_{\boldsymbol{\theta}_u}$, producing a deformation field $\boldsymbol{\phi}_i$ that is used to warp the template segmentation $\boldsymbol{S}^t$. The warped template segmentation $\boldsymbol{S}^t \circ \boldsymbol{\phi}_i$ is subsequently restored to the original resolution and used for full-thickness cartilage loss (FCL) estimation. Cartilage surface segmentation is followed by thickness mapping. A rule-based cartilage parcellation algorithm is developed for the regional quantification of cartilage thickness, FCL, surface area, and volume. The legend illustrates the utilization of training and testing/evaluation datasets.}
\label{fig:framework-overview}
\end{figure}

\subsection{Framework Overview}
\label{subsec:method-overview}
The objective of this work was to establish a pipeline for automated knee articular cartilage morphometrics. Fig. \ref{fig:framework-overview} shows the workflow of the proposed framework. It involves three deep learning models (\emph{i.e.}, $\mathcal{F}_{\boldsymbol{\theta}_s}$ for tissue segmentation, $\mathcal{G}_{\boldsymbol{\theta}_t}$ for template construction, and $\mathcal{G}_{\boldsymbol{\theta}_u}$ for image registration) in the upstream process. The downstream process includes cartilage thickness mapping, FCL estimation, cartilage parcellation, and regional quantification. Since deep learning models learn parameters from data during training and make inferences during testing, our framework complies with this two-stage strategy.

\subsubsection{Training Stage}
\label{subsubsec:method-overview-train}
Given the dataset $\{ \{\boldsymbol{I}_i, \boldsymbol{S}^l_i\} \}_{i=1}^n$ of size $n$ defined in the 3D image domain $\Omega \subset \mathbb{R}^3$, each image $\boldsymbol{I}_i$ is masked by segmentation label $\boldsymbol{S}^l_i$, downsampled via linear interpolation, and cropped into a low-resolution image: $\boldsymbol{I}^{\text{low}}_i = \mathcal{M}^\text{down}_{\boldsymbol{\theta}_i}(\boldsymbol{I}_i,\boldsymbol{S}^l_i)$. A knee template image $\boldsymbol{I}^t$ is learned from the template construction network $\mathcal{G}_{\boldsymbol{\theta}_t}$ trained with a subset of low-resolution images $\{ \boldsymbol{I}^{\text{low}}_i \}_{i=1}^{n_\text{t}}$ that comprises $n_\text{t}$ 3D images of subjects without knee OA. The template segmentation $\boldsymbol{S}^t$ is created via manual labeling or image-to-template registration. A network $\mathcal{F}_{\boldsymbol{\theta}_s}$ is trained for bone and cartilage segmentation at the original resolution with the training set $\{ \{\boldsymbol{I}_i, \boldsymbol{S}^l_i\} \}_{i=1}^{n_{\text{tr}}}$ of size $n_{\text{tr}}$. A registration network $\mathcal{G}_{\boldsymbol{\theta}_u}$ is trained on the low-resolution training images set $\{ \boldsymbol{I}^{\text{low}}_i \}_{i=1}^{n_{\text{tr}}}$ of size $n_{\text{tr}}$. We used low-resolution images during the training of $\mathcal{G}_{\boldsymbol{\theta}_t}$ and $\mathcal{G}_{\boldsymbol{\theta}_u}$ to reduce the demand for GPU memory.

\subsubsection{Inference Stage}
\label{subsubsec:method-overview-infer}
An input image $\boldsymbol{I}_i$ is segmented by the network $\mathcal{F}_{\boldsymbol{\theta}_s}$ which
outputs the respective bone and cartilage segmentation $\boldsymbol{S}_i$.
Sequentially, the template image $\boldsymbol{I}^t$ is registered to the
low-resolution image $\boldsymbol{I}^{\text{low}}_i = \mathcal{M}^\text{down}_{\boldsymbol{\theta}_i}(\boldsymbol{I}_i,\boldsymbol{S}_i)$, where a deformation field $\boldsymbol{\phi}_i$ is
estimated from the network $\mathcal{G}_{\boldsymbol{\theta}_u}$. The field $\boldsymbol{\phi}_i$ is used to warp
the template segmentation $\boldsymbol{S}^t$ so that it aligns with the low-resolution
image $\boldsymbol{I}^{\text{low}}_i$. The warped template segmentation $\boldsymbol{S}^t \circ \boldsymbol{\phi}_i$ is then
restored to the original resolution and space via a mapping function
$\mathcal{M}^\text{up}(\boldsymbol{S}^t \circ \boldsymbol{\phi}_i, {\boldsymbol{\theta}}_i)$ with parameters ${\boldsymbol{\theta}}_i$ estimated in the previous downsampling
step. Note that $\circ$ represents the spatial
transformation\cite{jaderberg2015spatial}. The abovementioned upstream
pipeline only involves evaluating the functions $\mathcal{F}_{\boldsymbol{\theta}_s}(\cdot)$ and $\mathcal{G}_{\boldsymbol{\theta}_u}(\cdot,\cdot)$
with learned parameters $\boldsymbol{\theta}_s$ and $\boldsymbol{\theta}_u$. Downstream tasks are
described in \ref{subsec:method-CTM} to \ref{subsec:method-regionalQuant}.

\subsection{Construction of Template Image \& Segmentation}
\label{subsec:method-template}

\begin{figure}[!t]
\centerline{\includegraphics[width=1\columnwidth]{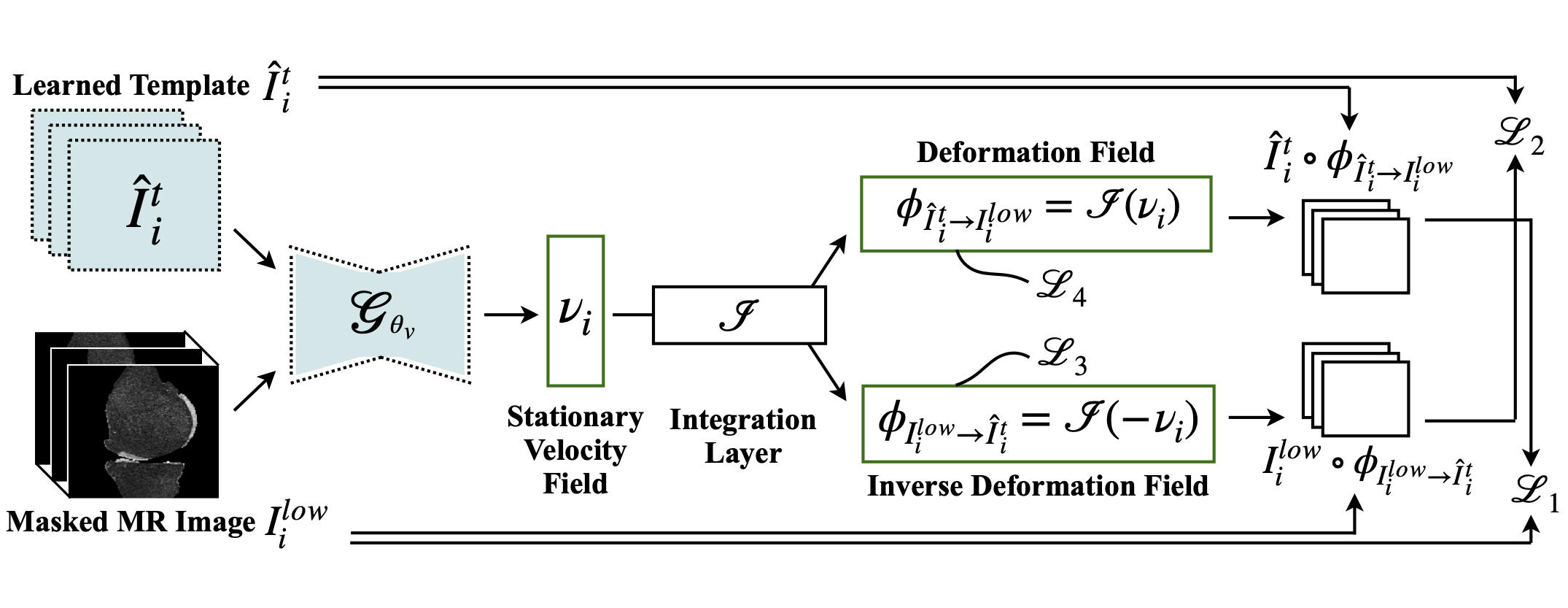}}
\caption{Model for template construction. 
A template image $\bm{I}^t$ is learned by minimizing the loss function $\mathcal{L}=\lambda_1\mathcal{L}_1+\lambda_2\mathcal{L}_2+\lambda_3\mathcal{L}_3+\lambda_4\mathcal{L}_4$, where $\{\lambda_1,\lambda_2,\lambda_3,\lambda_4\}$ are hyperparameters.}
\label{fig:model-template}
\end{figure}

We aimed to construct a representative template image that shows the normal
anatomy of bone and cartilage in a knee joint. We adopted a learning
model\cite{dalca2019learning} based on the stationary velocity field (SVF).
Related theories have been described in a probabilistic diffeomorphic
registration work\cite{dalca2019unsupervised}. Fig. \ref{fig:model-template}
shows the template construction network $\mathcal{G}_{\boldsymbol{\theta}_t}$ trained with low-resolution
images in $\{ \boldsymbol{I}^{\text{low}}_i \}_{i=1}^{n_\text{t}}$. We formulated the learning process as the optimization:
\begin{equation}
\label{eq:temp-1}
    \underset{\boldsymbol{\theta}_t}{\arg \, \min} \; \mathcal{L}(\boldsymbol{\theta}_t, \boldsymbol{I}^{\text{low}}_i).
\end{equation}
As shown in Fig. \ref{fig:model-template}, the network parameters $\boldsymbol{\theta}_t$ comprises a learnable template $\hat{\boldsymbol{I}^t_i}$ and the parameters $\boldsymbol{\theta}_v$ of a registration subnetwork $\mathcal{G}_{\boldsymbol{\theta}_v}$. The loss function in \ref{eq:temp-1} is thus written as
\begin{equation}
\label{eq:temp-2}
    \mathcal{L}(\boldsymbol{\theta}_t, \boldsymbol{I}^{\text{low}}_i) = \mathcal{L}(\boldsymbol{\theta}_v, \hat{\boldsymbol{I}^t_i}, \boldsymbol{I}^{\text{low}}_i) = \frac{1}{\mid \Omega\mid }\mathcal{L}_I(\boldsymbol{\theta}_v, \hat{\boldsymbol{I}^t_i}, \boldsymbol{I}^{\text{low}}_i),
\end{equation}
\begin{equation}
\label{eq:temp-3}
\begin{split}
    \mathcal{L}_I = 
    &\,  \lambda_1 \Vert \boldsymbol{I}^{\text{low}}_i - \hat{\boldsymbol{I}^t_i} \circ \mathcal{I}(\mathcal{G}_{\boldsymbol{\theta}_v}^i(\hat{\boldsymbol{I}^t_i},\boldsymbol{I}^{\text{low}}_i)) \Vert^2_2 \\
    &+  \lambda_2 \Vert \boldsymbol{I}^{\text{low}}_i \circ \mathcal{I}(-\mathcal{G}_{\boldsymbol{\theta}_v}^i (\hat{\boldsymbol{I}^t_i},\boldsymbol{I}^{\text{low}}_i)) - \hat{\boldsymbol{I}^t_i} \Vert^2_2 \\
    &+ \lambda_3 \Vert \mathcal{K}(\{ \mathcal{I}(-\mathcal{G}_{\boldsymbol{\theta}_v}^j (\hat{\boldsymbol{I}^t_j},\boldsymbol{I}^{\text{low}}_j)) \}_{j \in N(i)}) \Vert^2_2 \\
    &+  \lambda_4 \Vert \nabla \mathcal{I}(\mathcal{G}_{\boldsymbol{\theta}_v}^i(\hat{\boldsymbol{I}^t_i},\boldsymbol{I}^{\text{low}}_i)) \Vert^2_2,
\end{split}
\end{equation}

where $\mid \Omega\mid $ denotes the domain size (\emph{i.e.}, the total number of
voxels), $\mathcal{I}(\cdot)$ denotes numerical integration via scaling and
squaring\cite{arsigny2006log,dalca2019unsupervised,krebs2019learning},
$\mathcal{K}(\cdot)$ denotes weighted running average\cite[Eq. (6)]{dalca2019learning}, $N(i)$ denotes the $k$ predecessors of
$i$, and $\nabla$ is the gradient operator. The registration
subnetwork $\mathcal{G}_{\boldsymbol{\theta}_v}$ predicts an SVF which goes through the differentiable
integration layer $\mathcal{I}(\cdot)$ to produce a deformation field. As shown in Fig. \ref{fig:model-template}, we calculated the SVF as $\boldsymbol{\nu}_i = \mathcal{G}_{\boldsymbol{\theta}_v}^i(\hat{\boldsymbol{I}^t_i}, \boldsymbol{I}^{\text{low}}_i)$, the
template-to-image deformation field as $\boldsymbol{\phi}_{\hat{\boldsymbol{I}^t} \rightarrow \boldsymbol{I}^{\text{low}}_i} = \mathcal{I}(\boldsymbol{\nu}_i)$, and the image-to-template
deformation field as $\boldsymbol{\phi}_{\boldsymbol{I}^{\text{low}}_i \rightarrow \hat{\boldsymbol{I}^t}} = \mathcal{I}(-\boldsymbol{\nu}_i)$. Note that we enforced inverse-consistency
constraint inspired by previous studies\cite{zhang2018inverse,mok2020fast} in
the first two terms of \ref{eq:temp-3}. Specifically, the first term
encourages image similarity between the warped template $\hat{\boldsymbol{I}^t_i} \circ \mathcal{I}(\mathcal{G}_{\boldsymbol{\theta}_v}^i(\hat{\boldsymbol{I}^t_i},\boldsymbol{I}^{\text{low}}_i))$ and the
input image $\boldsymbol{I}^{\text{low}}_i$, whereas the second term encourages similarity between
the warped input image $\boldsymbol{I}^{\text{low}}_i \circ \mathcal{I}(-\mathcal{G}_{\boldsymbol{\theta}_v}^i (\hat{\boldsymbol{I}^t_i},\boldsymbol{I}^{\text{low}}_i))$ and the template $\hat{\boldsymbol{I}^t_i}$. The third term
restricts the magnitude of deformation fields and the fourth term encourages
smoothness of deformation fields. Consequently, a template image can be learned
from the network: $\boldsymbol{I}^t=\hat{\boldsymbol{I}^{t*}_i}$, where $*$ denotes optimal estimation.
Finally, the template segmentation is constructed by manual labeling or
image-to-template registration. The registration approach relies on the
calculation of a probability map for each ROI, which is formulated as
\begin{equation}
\label{eq:temp-4}
     \frac{1}{n_\text{t}} \sum_{i=1}^{n_\text{t}}  (\boldsymbol{S}_i^{\text{low}-l} \circ \mathcal{I}(-\mathcal{G}_{\boldsymbol{\theta}_v}(\boldsymbol{I}^t, \boldsymbol{I}_i^{\text{low}}))),
\end{equation}
where $\{\boldsymbol{S}_i^{\text{low}-l}, \boldsymbol{I}_i^{\text{low}}\}_{i=1}^{n_\text{t}} = \mathcal{M}^\text{down}_{\{ \boldsymbol{\theta}_{i} \}} (\{ \boldsymbol{I}_i \}, \{ \boldsymbol{S}^{l}_i \})$.
The implementation details are given in  \ref{subsec:experiment-1-template}.

\subsection{Cartilage \& Bone Segmentation}
\label{subsec:method-segNet}
In this part, we describe the segmentation model but emphasize that it is
flexible to change to a wide variety of deep learning models owing to the
modular design of the proposed framework. The nnU-Net\cite{isensee2021nnu}
was adopted for not only its competitive segmentation accuracy but also its
self-configuring characteristic and superior generalizability. We trained
several variants of the nnU-Net model and chose the best for inference.
Experimental details are given in  \ref{subsec:experiment-2-segNet}.

With the training set $\{ \{\boldsymbol{I}_i, \boldsymbol{S}^l_i \}\}_{i=1}^{n_{\text{tr}}}$, we formulated the model training process as learning a parameterized function $\mathcal{F}_{\boldsymbol{\theta}_s}(\boldsymbol{I}_i)=\boldsymbol{S}_i$ that takes an image $\boldsymbol{I}_i$ as input and outputs respective tissue segmentation $\boldsymbol{S}_i$. The model was optimized by minimizing the loss function $\mathcal{L}(\mathcal{F}_{\boldsymbol{\theta}_s}(\boldsymbol{I}_i), \boldsymbol{S}^l_i)$ that quantifies the dissimilarity between model segmentation $\boldsymbol{S}_i$ and manual labels $\boldsymbol{S}^l_i$. Note that the nnU-Net framework automates a postprocessing procedure. Specifically, connectivity-based postprocessing is applied to the merged mask and individual masks, and a combination of postprocessing steps that improve the segmentation performance is retained. The nnU-Net model is therefore formulated as $\mathcal{P}(\mathcal{F}_{\boldsymbol{\theta}_s}(\boldsymbol{I}_i))=\boldsymbol{S}_i$ where $\mathcal{P}(\cdot)$ denotes postprocessing.

\subsection{Template-to-image Registration}
\label{subsec:method-regNet}

\begin{figure}[!t]
\centerline{\includegraphics[width=1\columnwidth]{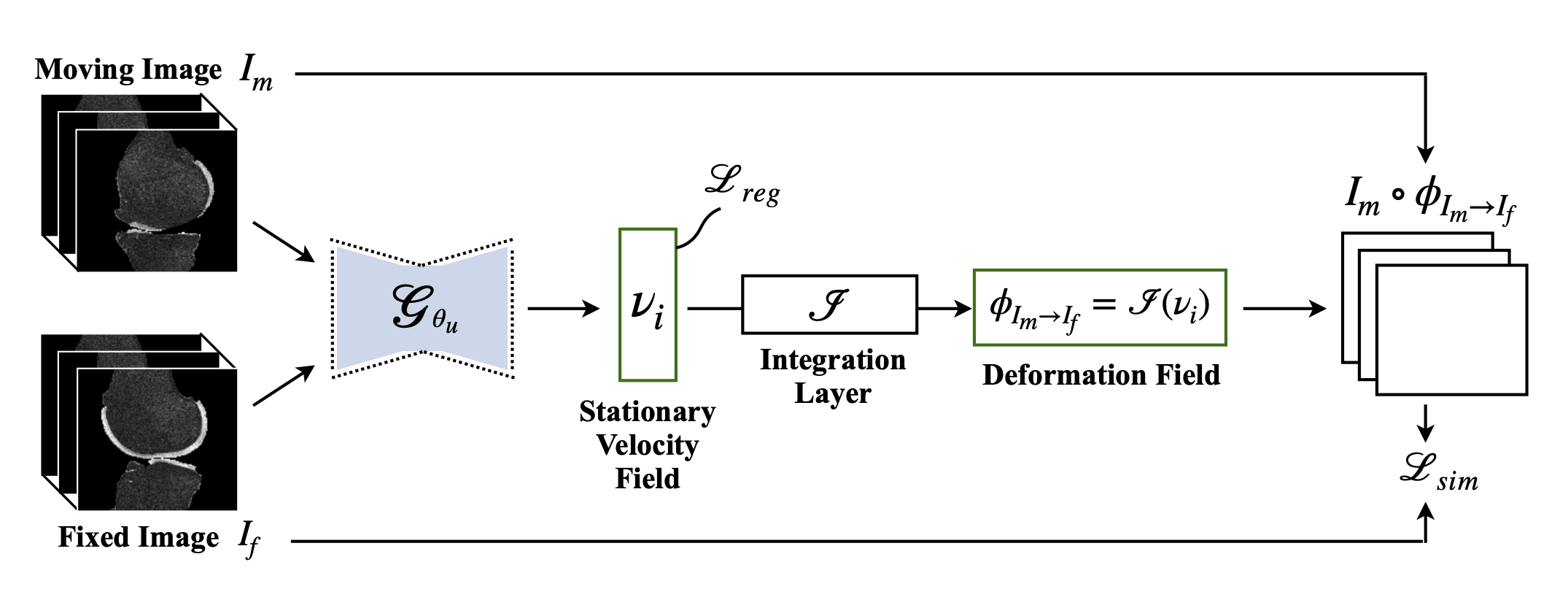}}
\caption{Model for image registration. 
A global parameterized function $\mathcal{G}_{\bm{\theta}_u}(\bm{I}_{m}, \bm{I}_{f})$ is learned by minimizing the loss function $\mathcal{L}=\mathcal{L}_\text{sim}+\lambda \mathcal{L}_\text{reg}$, where $\lambda$ is hyperparameter.}
\label{fig:model-reg}
\end{figure}

The purpose of training a registration model is to use the anatomical prior
knowledge in the template segmentation $\boldsymbol{S}^t$ for FCL estimation by
registering the template image $\boldsymbol{I}^t$ to the low-resolution image
$\boldsymbol{I}^{\text{low}}_i$. The model was trained with random pairs of images in the
low-resolution training set $\{ \boldsymbol{I}^{\text{low}}_i \}_{i=1}^{n_{\text{tr}}}$ and used for template-to-image
registration. We adopted an unsupervised deformable registration model called
VoxelMorph\cite{balakrishnan2019voxelmorph}. Let a random pair of moving and
fixed images be $\{ \boldsymbol{I}_m, \boldsymbol{I}_f \}$. The registration task is modeled as learning a
global parameterized function $\mathcal{G}_{\boldsymbol{\theta}_u}(\boldsymbol{I}_{m}, \boldsymbol{I}_{f})$ that estimates a deformation field
$\boldsymbol{\phi}_{\boldsymbol{I}_m \rightarrow \boldsymbol{I}_f}$ for aligning the moving image $\boldsymbol{I}_{m}$ to the fixed image
$\boldsymbol{I}_{f}$ (Fig. \ref{fig:model-reg}). The learning process is formulated as the
optimization:
\begin{equation}
\label{eq:reg-1}
    \underset{\boldsymbol{\theta}_u}{\arg \, \min} \; \mathcal{L}(\boldsymbol{\theta}_u, \boldsymbol{I}_m, \boldsymbol{I}_f),
\end{equation}
\begin{equation}
\label{eq:reg-2}
    \mathcal{L}(\boldsymbol{\theta}_u, \boldsymbol{I}_m, \boldsymbol{I}_f) = \mathcal{L}_{\text{sim}}(\boldsymbol{\theta}_u, \boldsymbol{I}_m, \boldsymbol{I}_f) + \lambda \mathcal{L}_{\text{reg}}(\boldsymbol{\theta}_u, \boldsymbol{I}_m, \boldsymbol{I}_f), 
\end{equation}
where $\mathcal{L}_{\text{sim}}(\cdot,\cdot,\cdot)$ quantifies image dissimilarity and $\mathcal{L}_{\text{reg}}(\cdot,\cdot,\cdot)$ is the
regularization term. Some commonly used image similarity losses include mean
squared error (MSE), cross-correlation (CC), normalized cross-correlation
(NCC), local normalized cross-correlation (LNCC), mutual information (MI), and
modality independent neighborhood descriptor (MIND)\cite{heinrich2012mind}.
We compared MSE and LNCC in this work. The MSE is calculated as
\begin{equation}
\label{eq:reg-3}
\begin{split}
    \mathcal{L}_{\text{sim}}(\boldsymbol{\theta}_u, \boldsymbol{I}_m, \boldsymbol{I}_f) 
    &= \mathcal{L}_{\text{MSE}}(\boldsymbol{\theta}_u, \boldsymbol{I}_m, \boldsymbol{I}_f) \\
    &=  \frac{1}{\mid \Omega\mid } \Vert \boldsymbol{I}_m \circ \mathcal{I}(\mathcal{G}_{\boldsymbol{\theta}_u}(\boldsymbol{I}_m, \boldsymbol{I}_f)) - \boldsymbol{I}_f \Vert^2_2.
\end{split}
\end{equation}
Let $\Tilde{\boldsymbol{I}}_m = \boldsymbol{I}_m \circ \mathcal{I}(\mathcal{G}_{\boldsymbol{\theta}_u}(\boldsymbol{I}_m, \boldsymbol{I}_f))$ be the warped moving image, then the LNCC is calculated as
\begin{equation}
\label{eq:reg-4}
\begin{split}
    \mathcal{L}_{\text{sim}}(\boldsymbol{\theta}_u, \boldsymbol{I}_m, \boldsymbol{I}_f) 
    &= \mathcal{L}_{\text{LNCC}}(\boldsymbol{\theta}_u, \boldsymbol{I}_m, \boldsymbol{I}_f) \\
    &= -\frac{1}{\mid \Omega\mid }\sum_{p \in \Omega} \mathcal{C}(\Tilde{\boldsymbol{I}}_m(p), \boldsymbol{I}_f(p)),
\end{split}
\end{equation}
\begin{equation}
\label{eq:reg-5}
\begin{split}
    &\mathcal{C}(\Tilde{\boldsymbol{I}}_m(p), \boldsymbol{I}_f(p)) = \\
    &\frac{
    \left[ \sum_{q \in N(p)} (\Tilde{\boldsymbol{I}}_m(q) - \bar{\Tilde{\boldsymbol{I}}}_m^p) (\boldsymbol{I}_f(q) - \bar{\boldsymbol{I}}_f^p) \right]^2
    }
    {
     \left[\sum_{q \in N(p)}(\Tilde{\boldsymbol{I}}_m(q) - \bar{\Tilde{\boldsymbol{I}}}_m^p)^2 \right] \left[\sum_{q \in N(p)}(\boldsymbol{I}_f(q) - \bar{\boldsymbol{I}}_f^p)^2 \right]
    },
\end{split}
\end{equation}
where $\mathcal{C}(\cdot, \cdot)$ calculates the LNCC, $N(p)$ is the
$k$-neighborhood of $p$,  $\bar{\Tilde{\boldsymbol{I}}}_m^p$ and $\bar{\boldsymbol{I}}_f^p$ are
the averages of a local volume centered around $p$ in $\Tilde{\boldsymbol{I}}_m$ and
$\boldsymbol{I}_f$, respectively. In contrast to previous
work\cite{balakrishnan2019voxelmorph} in which the deformation field
$\boldsymbol{\phi}_{\boldsymbol{I}_m \rightarrow \boldsymbol{I}_f}$ was directly estimated, we predicted the SVF from the registration
network $\mathcal{G}_{\boldsymbol{\theta}_u}$ and calculated the deformation field via the
differentiable integration layer $\mathcal{I}(\cdot)$: $\boldsymbol{\phi}_{\boldsymbol{I}_m \rightarrow \boldsymbol{I}_f} = \mathcal{I}(\mathcal{G}_{\boldsymbol{\theta}_u}(\boldsymbol{I}_{m}, \boldsymbol{I}_{f}))$. In this work, we
enforced smoothness constraints on the SVF:
\begin{equation}
\label{eq:reg-6}
\mathcal{L}_{\text{reg}} = \; \frac{1}{\mid \Omega\mid } \Vert \nabla \mathcal{G}_{\boldsymbol{\theta}_u}(\boldsymbol{I}_m,\boldsymbol{I}_f) \Vert^2_2.
\end{equation}
The implementation details are given in  \ref{subsec:experiment-3-regNet}.

\subsection{Cartilage Thickness Mapping}
\label{subsec:method-CTM}

We established an automated image processing pipeline for cartilage thickness mapping. The cartilage surface is reconstructed from the respective segmentation masks. The bone-cartilage interface is extracted via the proposed surface segmentation algorithm. The cartilage thickness is assessed via a surface-normal-based thickness mapping algorithm.

\subsubsection{Cartilage Surface Segmentation}
\label{subsubsec:method-CTM-surfSeg}

Given the cartilage mask $\boldsymbol{S}_{i,c}$ and corresponding bone mask $\boldsymbol{S}_{i,b}$ of an image $\boldsymbol{I}_i$, voxels on the bone--cartilage interface (\emph{i.e.}, inner surface) are extracted according to
\begin{equation}
\label{eq:CTM-1}
    \boldsymbol{V}_{i,c}^{\text{in}} = \mathcal{O}_{\text{boundary}}(\boldsymbol{S}_{i,c}) \cap \mathcal{O}_{\text{boundary}}(\mathcal{O}_{\text{inverse}}(\mathcal{O}_{\text{filling}}(\boldsymbol{S}_{i,c}, \boldsymbol{S}_{i,b}) - \boldsymbol{S}_{i,c})),
\end{equation}
where $\mathcal{O}_{\text{boundary}}(\cdot)$ denotes the boundary extractor, $\mathcal{O}_{\text{inverse}}(\cdot)$ denotes the inverse operation, and $\mathcal{O}_{\text{filling}}(\cdot,\cdot)$ denotes gap filling via bone region growing. Voxels on the outer surface of cartilage are extracted according to
\begin{equation}
\label{eq:CTM-2}
    \boldsymbol{V}_{i,c}^{\text{out}} = \mathcal{O}_{\text{boundary}}(\boldsymbol{S}_{i,c}) - \boldsymbol{V}_{i,c}^{\text{in}}.
\end{equation}

\begin{figure}[!t]
\centerline{\includegraphics[width=0.8\columnwidth]{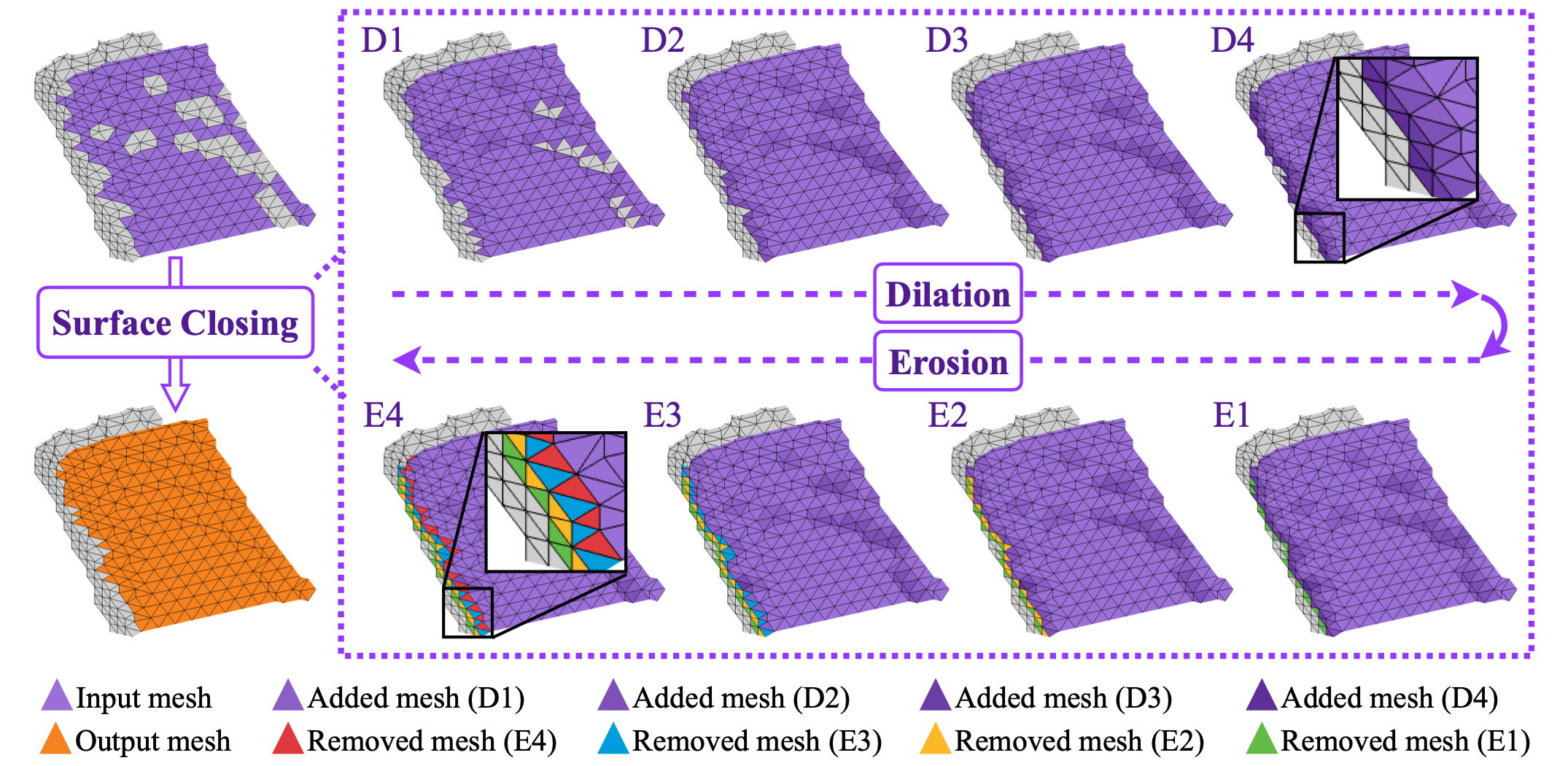}}
\caption{Example results of a surface closing with 4 surface dilation and 4 surface erosion operations. One layer of meshes is added to the surface edge in a surface dilation operation, whereas marginal meshes are removed in a surface erosion operation.}
\label{fig:surfClosing}
\end{figure}

To improve the accuracy and robustness of surface segmentation, we fine-tuned the coarse results from \ref{eq:CTM-1} using the proposed surface closing operation (Fig. \ref{fig:surfClosing}) and further adjusted the results from \ref{eq:CTM-2} through a proposed restricted surface dilation. Let the mesh representation of the cartilage and the inner  surface be $\boldsymbol{M}_{i,c}$ and $\boldsymbol{M}_{i,c}^{\text{in}}$, respectively. The proposed surface closing operation $\mathcal{O}_{c}(\cdot)$ is defined in the surface domain $\Omega_{\text{surf}} \subset \mathbb{R}^2$. The surface closing operation applied to the inner surface $\boldsymbol{M}_{i,c}^{\text{in}}$ is formulated as
\begin{equation}
\label{eq:CTM-3}
    \mathcal{O}_{\text{closing}}(\boldsymbol{M}_{i,c}^{\text{in}} \mid  \Omega_{\text{surf}}) = \mathcal{O}_{\text{erosion}}^{n_\text{e}}( \mathcal{O}_{\text{dilation}}^{n_\text{d}}(\boldsymbol{M}_{i,c}^{\text{in}} \mid  \Omega_{\text{surf}}) ),
\end{equation}
where $\mathcal{O}_{\text{erosion}}^{n_\text{e}}(\cdot)$ denotes compound surface erosion with $n_\text{e}$ iterations and $\mathcal{O}_{\text{dilation}}^{n_\text{d}}(\cdot \vert  \; \Omega_{\text{surf}})$ denotes compound surface dilation in a surface domain $\Omega_{\text{surf}}$ with $n_\text{d}$ iterations. Note that the proposed surface dilation adds one layer of mesh to the edge, whereas surface erosion removes the peripheral mesh. In this work, the surface closing operation is restricted in the cartilage mesh: $\Omega_{\text{surf}} = \boldsymbol{M}_{i,c}$. 

The proposed restricted surface dilation $\mathcal{O}_{\text{r-dilation}}(\cdot)$ involves iterative surface dilations with a boundary restriction $\mathbb{B}$. It is defined in the restricted surface domain $\Omega_{\text{surf}}^{\mathbb{B}} \subset \Omega_{\text{surf}}$. Let the mesh representation of the outer surface be $\boldsymbol{M}_{i,c}^{\text{out}}$. In this work, we applied restricted surface dilation $\mathcal{O}_{\text{r-dilation}}(\cdot)$ to the outer surface $\boldsymbol{M}_{i,c}^{\text{out}}$ with the restricted surface domain defined as $\Omega_{\text{surf}}^{\mathbb{B}} = \boldsymbol{M}_{i,c} - \mathcal{O}_{\text{closing}}(\boldsymbol{M}_{i,c}^{\text{in}} \mid  \Omega_{\text{surf}})$. The fine-tuning of the outer surface is thus formulated as
\begin{equation}
\label{eq:CTM-4}
    \mathcal{O}_{\text{r-dilation}}(\boldsymbol{M}_{i,c}^{\text{out}} \mid  \Omega_{\text{surf}}, \mathbb{B}) = \mathcal{O}_{\text{dilation}}^\infty(\boldsymbol{M}_{i,c}^{\text{out}} \mid  \Omega_{\text{surf}}^{\mathbb{B}}),
\end{equation}
where $\mathcal{O}_{\text{dilation}}^\infty(\cdot \vert  \; \Omega_{\text{surf}}^{\mathbb{B}})$ denotes iterative surface dilations in a restricted surface
domain $\Omega_{\text{surf}}^{\mathbb{B}}$ until no mesh can be added.

\subsubsection{Cartilage Thickness Measurement}
\label{subsubsec:method-CTM-thickness}

The proposed surface-normal-based thickness mapping approach comprises surface normal estimation, orientation correction, spatial smoothing, and thickness measurement. The surface normal for each voxel on the bone--cartilage interface $\boldsymbol{M}_{i,c}^{\text{in}}$ is estimated through singular value decomposition (SVD). Let $\mathfrak{L}(\cdot)$ be a locator function that returns the position of the input voxel in the image domain $\Omega$. A position matrix $\boldsymbol{P}$ is then constructed for each voxel $p \in \boldsymbol{M}_{i,c}^{\text{in}}$ so that
\begin{equation}
\label{eq:CTM-5}
    \boldsymbol{P} = \tilde{\boldsymbol{P}} - \bar{\boldsymbol{P}},
\end{equation}
\begin{equation}
\label{eq:CTM-6}
    \tilde{\boldsymbol{P}}(j,:) = \mathfrak{L}(q), \quad \forall q \in N(p), \; j=1, \ldots ,\mid N(p)\mid ,
\end{equation}
\begin{equation}
\label{eq:CTM-7}
    \bar{\boldsymbol{P}} = \frac{1}{\mid N(p)\mid } {\mathbbm{1}} \tilde{\boldsymbol{P}}, \quad {\mathbbm{1}} \in \mathbb{R}^{1,\mid N(p)\mid },
\end{equation}
where $N(p)$ denotes the $k$-nearest-neighborhood of $p$, $\mid N(p)\mid $ denotes the neighborhood size $k$, $\bar{\boldsymbol{P}}$ denotes the row-wise average of $\tilde{\boldsymbol{P}}$, and $\tilde{\boldsymbol{P}} \in \mathbb{R}^{k,3}$. The matrix $\boldsymbol{P}$ can be decomposed into the form $\boldsymbol{U} \boldsymbol{\sum} \boldsymbol{V}^\top$ via SVD, where $\boldsymbol{U}$ and $\boldsymbol{V}$ are orthogonal matrices, and $\boldsymbol{\sum}$ is a diagonal matrix with non-zero diagonal entries called singular values $\boldsymbol{\sigma}_{i,i}=\boldsymbol{\sum}_{i,i}$. The columns of $\boldsymbol{U}$ and $\boldsymbol{V}$ are referred to as the left-singular vectors and right-singular vectors of $\boldsymbol{P}$, respectively. Following our construction of matrix $\boldsymbol{P}$ in \ref{eq:CTM-5}--\ref{eq:CTM-7}, the right-singular vector in $\boldsymbol{V}$ that corresponds to the minimal singular value in $\boldsymbol{\sum}$ represents the direction of least spatial variation. With the matrix $\boldsymbol{P}$, the surface normal at location $\mathfrak{L}(p)$ is estimated by
\begin{equation}
\label{eq:CTM-8}
    \tilde{\boldsymbol{N}}(p) = \mathcal{O}_{\text{SVD}}(\mathcal{S}(\boldsymbol{P})), \quad p \in \boldsymbol{M}_{i,c}^{\text{in}},
\end{equation}
where $\tilde{\boldsymbol{N}}(p)$ denotes the initial estimation of the surface normal at $\mathfrak{L}(p)$, $\mathcal{S}(\cdot)$ denotes SVD, and $\mathcal{O}_{\text{SVD}}(\cdot)$ denotes the selection of the right-singular vector. The selected right-singular vector only captures the direction with the least data variation, and the estimated surface normals thus point arbitrarily inwards and outwards with regard to the cartilage region. An orientation correction algorithm was developed to make the surface normals point inwards.  A spatial smoothing algorithm was developed to improve the spatial consistency of the estimated surface normals. Consequently, the estimated surface normal at $\mathfrak{L}(p)$ is written as
\begin{equation}
\label{eq:CTM-9}
    \hat{\boldsymbol{N}}(p) = \mathcal{O}_{\text{smoothing}}(\mathcal{O}_{\text{reorientation}}(\tilde{\boldsymbol{N}}(p))), \quad p \in \boldsymbol{M}_{i,c}^{\text{in}},
\end{equation}
where $\mathcal{O}_{\text{reorientation}}(\cdot)$ denotes the orientation correction and $\mathcal{O}_{\text{smoothing}}(\cdot)$ denotes the spatial smoothing. With the estimated surface normals $\hat{\boldsymbol{N}}$, cartilage thicknesses $\hat{\boldsymbol{T}}$ can be measured along the normal directions:
\begin{equation}
\label{eq:CTM-10}
    \hat{\boldsymbol{T}} = \mathcal{O}_{\text{thickness}}(\boldsymbol{M}_{i,c}^{\text{in}}, \boldsymbol{M}_{i,c}^{\text{out}}, \hat{\boldsymbol{N}}),
\end{equation}
where $\mathcal{O}_{\text{thickness}}(\cdot)$ denotes the surface-normal-based thickness measurement. 

\begin{figure}[!t]
\centerline{\includegraphics[width=0.8\columnwidth]{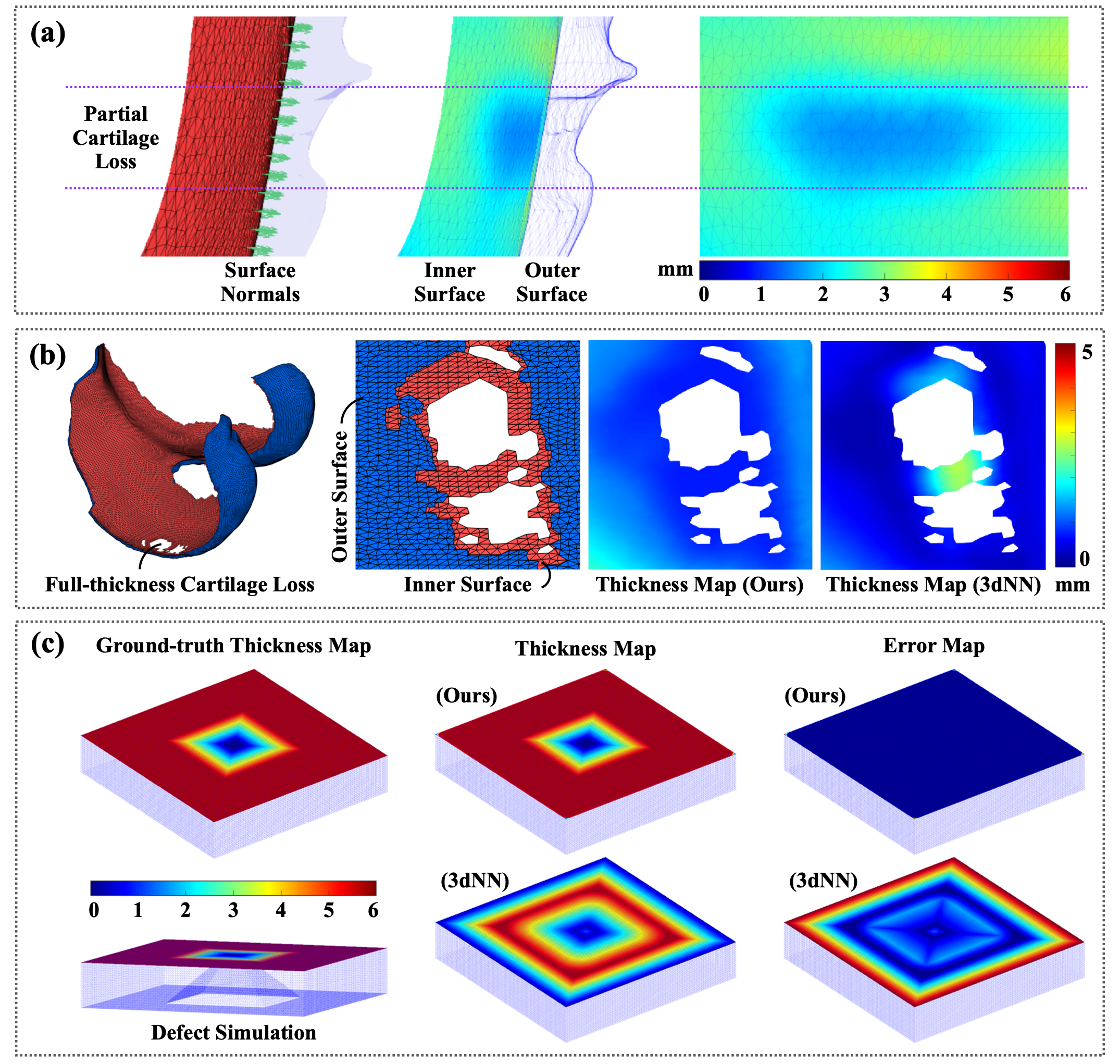}}
\caption{Example results of the proposed surface-normal-based thickness mapping method. (a) Partial cartilage loss can be  reflected in the thickness map obtained from our method. (b) Compared with the 3D nearest neighbor (3dNN) approach, our thickness mapping method is more accurate in peripheral regions and regions with thin cartilage. (c) The thickness maps of synthetic data with defect simulation obtained from our method and the 3dNN method are visualized.}
\label{fig:CTMExample}
\end{figure}

Fig. \ref{fig:CTMExample}a shows that partial cartilage loss can be reflected on the cartilage thickness map. This is attributed to the accurate estimation of surface normals. Additionally, the spatial smoothing of surface normals improves the smoothness of the thickness map. Fig. \ref{fig:CTMExample}b shows the superiority of the surface-normal-based method over the 3D nearest neighbor (3dNN) approach. In cartilaginous regions with one layer of voxels, the 3dNN method shows overestimation in central areas and underestimation in areas where the inner and outer surfaces are in contact. The superiority of the purposed method is further validated with synthetic data (Fig. \ref{fig:CTMExample}c). A cuboid with defect simulation was synthesized, and the respective ground-truth thickness map was obtained. The thickness maps and error maps indicate that our method is robust to cartilage defects and more accurate in peripheral regions.

\begin{figure}[!t]
\centerline{\includegraphics[width=1\columnwidth]{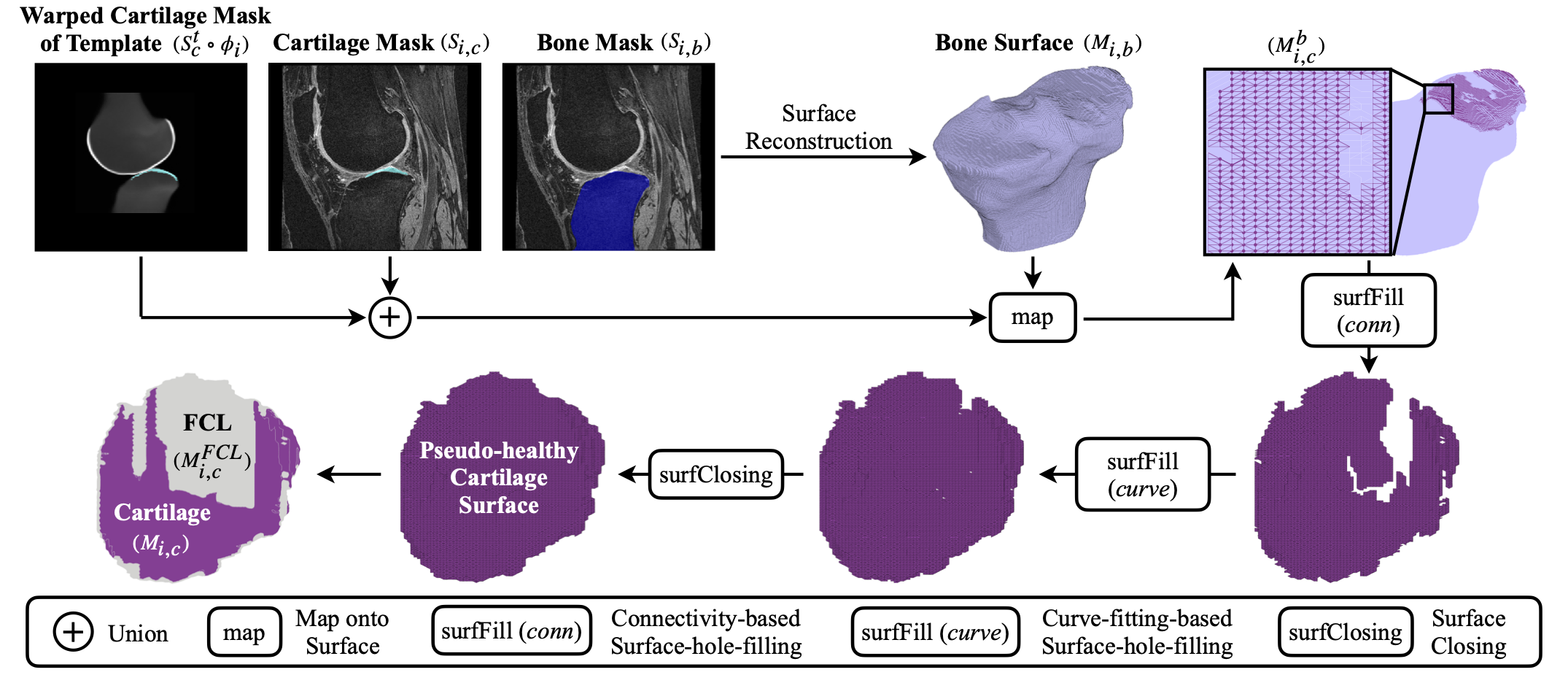}}
\caption{Pipeline of the proposed full-thickness cartilage loss (FCL) estimation method. The warped cartilage mask of the template image $\bm{S}^t_c \circ \bm{\phi}_i$ is merged with the cartilage mask $\bm{S}_{i,c}$ and mapped onto the reconstructed bone surface $\bm{M}_{i,b}$. Connectivity-based and curve-fitting-based surface-hole-filling methods are proposed to reconstruct a pseudo-healthy cartilage surface. Surface closing is applied to fine-tune the reconstructed surface. The FCL surface $\bm{M}^\text{FCL}_{i,c}$ is estimated by subtracting the cartilage surface $\bm{M}_{i,c}$ from the pseudo-healthy surface.}
\label{fig:FCL-estimation}
\end{figure}

\subsection{Full-thickness Cartilage Loss Estimation}
\label{subsec:method-FCL}

We calculated FCL as the percentage of the subchondral bone area without
cartilage coverage, which is equivalent to the percentage of the denuded area
(\emph{i.e.}, dABp)\cite[Tab. II]{eckstein2006proposal}. As shown in Fig. \ref{fig:FCL-estimation}, the FCL estimation module takes the cartilage mask
$\boldsymbol{S}_{i,c}$, bone mask $\boldsymbol{S}_{i,b}$, and warped cartilage mask of the template
$\boldsymbol{S}^t_c \circ \boldsymbol{\phi}_i$ as inputs. The cartilage masks $\boldsymbol{S}_{i,c}$ and $\boldsymbol{S}^t_{c} \circ \boldsymbol{\phi}_i$ are
first merged and mapped onto the surface $\boldsymbol{M}_{i,b}$ reconstructed from the
bone mask $\boldsymbol{S}_{i,b}$. Surface-hole-filling methods are proposed for the
reconstruction of a pseudo-healthy cartilage surface. The reconstructed surface
is further fine-tuned via surface closing \ref{eq:CTM-3}. Finally, the FCL
surface $\boldsymbol{M}_{i,c}^{\text{FCL}}$ is estimated by subtracting the cartilage surface
$\boldsymbol{M}_{i,c}$ from the pseudo-healthy surface. As shown in Fig. \ref{fig:FCL-estimation}, the FCL estimation method is formulated as
\begin{equation}
\label{eq:FCL-1}
    \boldsymbol{M}_{i,c}^{\text{FCL}} = \mathcal{O}_{\text{closing}}(\mathcal{O}_{\text{surface-filling}}(\boldsymbol{M}_{i,c}^b, \boldsymbol{M}_{i,b}) \mid  \boldsymbol{M}_{i,b}) - \boldsymbol{M}_{i,c},
\end{equation}
\begin{equation}
\label{eq:FCL-2}
    \boldsymbol{M}_{i,c}^b = \mathcal{O}_{\text{mapping}}(\boldsymbol{S}_{i,c} \cap (\boldsymbol{S}^t_{c} \circ \boldsymbol{\phi}_i), \boldsymbol{M}_{i,b}),
\end{equation}
where $\mathcal{O}_{\text{mapping}}(\cdot, 
\cdot)$ denotes voxel-to-surface mapping and $\mathcal{O}_{\text{surface-filling}}(\cdot,\cdot)$ denotes surface-hole-filling.

More specifically, the proposed surface-hole-filling  comprises connectivity-based and curve-fitting-based methods. After the mapping $\mathcal{O}_{\text{mapping}}(\cdot,\cdot)$, regions in the bone surface $\boldsymbol{M}_{i,b}$ enclosed by the cartilage surface $\boldsymbol{M}_{i,c}^b$ can be identified and filled through connectivity-based surface-hole-filling. The curve-fitting-based method can detect penetrative regions that extend to the edge of the cartilage. Note that the curve-fitting for the tibial cartilage is in the Cartesian coordinate system, whereas that for the femoral cartilage is in the polar coordinate system owing to its curved surface. The application of these two complementary surface-hole-filling strategies improves the robustness of FCL estimation.

\begin{figure}[!t]
\centerline{\includegraphics[width=0.8\columnwidth]{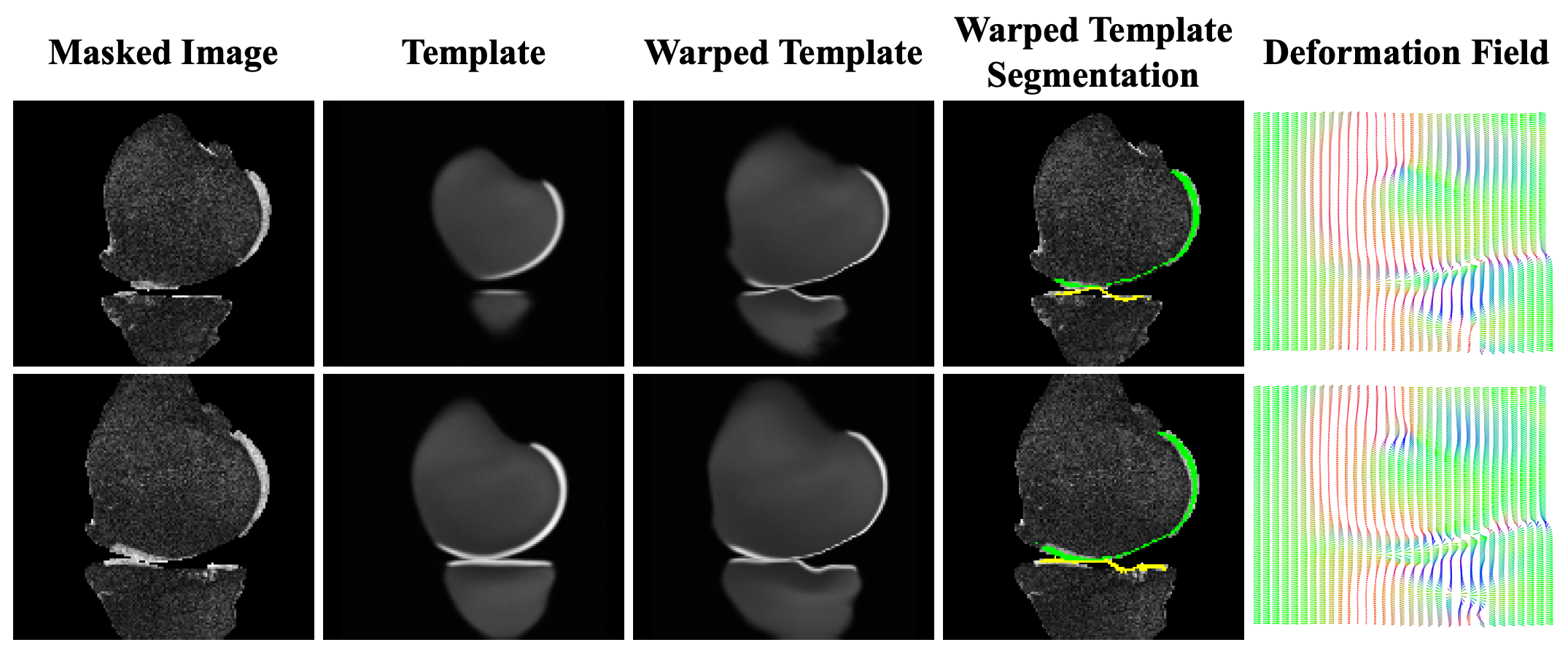}}
\caption{The effectiveness of template-to-image registration in full-thickness cartilage loss (FCL) estimation. The template is registered to the masked image, producing a deformation field that is used to warp the template segmentation. The warped template segmentation overlaid on the masked image shows that FCL can be reconstructed with the help of registration.}
\label{fig:regExample}
\end{figure}

In our method, prior knowledge of knee anatomy facilitates the FCL estimation. The segmentation masks of the template image represent the normal shapes and positions of bone and cartilage. Fig. \ref{fig:regExample} demonstrates the effectiveness of template-to-image registration in FCL estimation. It shows that the warped cartilage segmentation mask of the template covers not only cartilaginous regions but also denuded areas. Therefore, a coarse estimation of the pseudo-healthy cartilage surface can be obtained from the template-to-image registration. The proposed surface-based algorithms can further improve the accuracy and robustness of the FCL estimation.

\subsection{Rule-based Cartilage Parcellation}
\label{subsec:method-parcellation}

An automated rule-based cartilage parcellation algorithm was developed for
partitioning the femoral cartilage (FC), medial tibial cartilage (MTC), and
lateral tibial cartilage (LTC) into 20 regions as defined
previously\cite[Fig. 1]{wirth2009regional}. Compared with other parcellation
schemes\cite[Fig. 1]{favre2017anatomically}, the adopted scheme focuses more
on the weight-bearing regions of cartilage. 

Before cartilage parcellation, the MR image is reoriented according to the RAS+ convention where the positive axes of the image coordinate system point in the right, anterior, and superior directions. With the knee side information and consistent image orientation, the medial, lateral, interior, and exterior compartments can be identified automatically. The proposed method can be applied to either the cartilage surface or volume. In this work, the rule-based algorithm was applied to the pseudo-healthy cartilage surface. By adopting this method, the parcellation algorithm can demonstrate resilience against cartilage lesions.

The parcellation of FC is summarized as follows. First, the intercondylar notch is identified. A sagittal cutting plane that goes through the intercondylar notch splits the FC into medial and lateral femoral condyles (\emph{i.e.}, MFC \& LFC). The central regions of the femoral condyles (\emph{i.e.}, cMFC \& cLFC) are then defined as the areas between the anterior and posterior cutting planes. The anterior cutting plane goes through the intercondylar notch. Its posterior counterpart is located at a position corresponding to 60\% of the distance from the intercondylar notch to the posterior end of the femoral condyles. Both cutting planes are parallel to coronal slices. Additionally, these planes cut out the anterior and posterior compartments (\emph{i.e.}, aMFC, pMFC, aLFC, and pLFC). Finally, each central femoral condyle is split into 3 equal areas in the left--right direction. The exterior compartments (\emph{i.e.}, ecMFC \& ecLFC), central compartments (\emph{i.e.}, ccMFC \& ccLFC), and interior compartments (\emph{i.e.}, icMFC \& icLFC) are consequently defined.

The parcellation of tibial cartilage is summarized as follows. First, the
central region of the MTC or LTC (\emph{i.e.}, cMTC \& cLTC) is defined as an
elliptical ROI that covers 20\% of the cartilaginous surface. More
specifically, the ellipse is centered and aligned with the two principal
directions estimated from SVD. The estimation of principal directions is
similar to the surface normal estimation in
\ref{subsubsec:method-CTM-thickness} except that the right-singular vectors
corresponding to the two largest singular values are taken. Note that all
points on the surface of the MTC or LTC are used in SVD. Let $d_1$ be
the length of the semi-major axis of the ellipse, $d_2$ be the length of
the semi-minor axis, $\sigma_1$ be the largest singular value, and
$\sigma_2$ be the secondary singular value. The shape of the ellipse is
constrained by the equation $d_1/d_2=(\sigma_1/\sigma_2)^{1/2}$. The peripheral region of the tibial
cartilage is then subdivided into 4 compartments as described
previously\cite{wirth2008technique}. Finally, these compartments are labeled
as the anterior (\emph{i.e.}, aMTC \& aLTC), exterior (\emph{i.e.}, eMTC \&
eLTC), posterior (\emph{i.e.}, pMTC \& pLTC), and interior (\emph{i.e.}, iMTC
\& iLTC) parts according to their positions. 

\begin{figure}[!ht]
\centerline{\includegraphics[width=0.8\columnwidth]{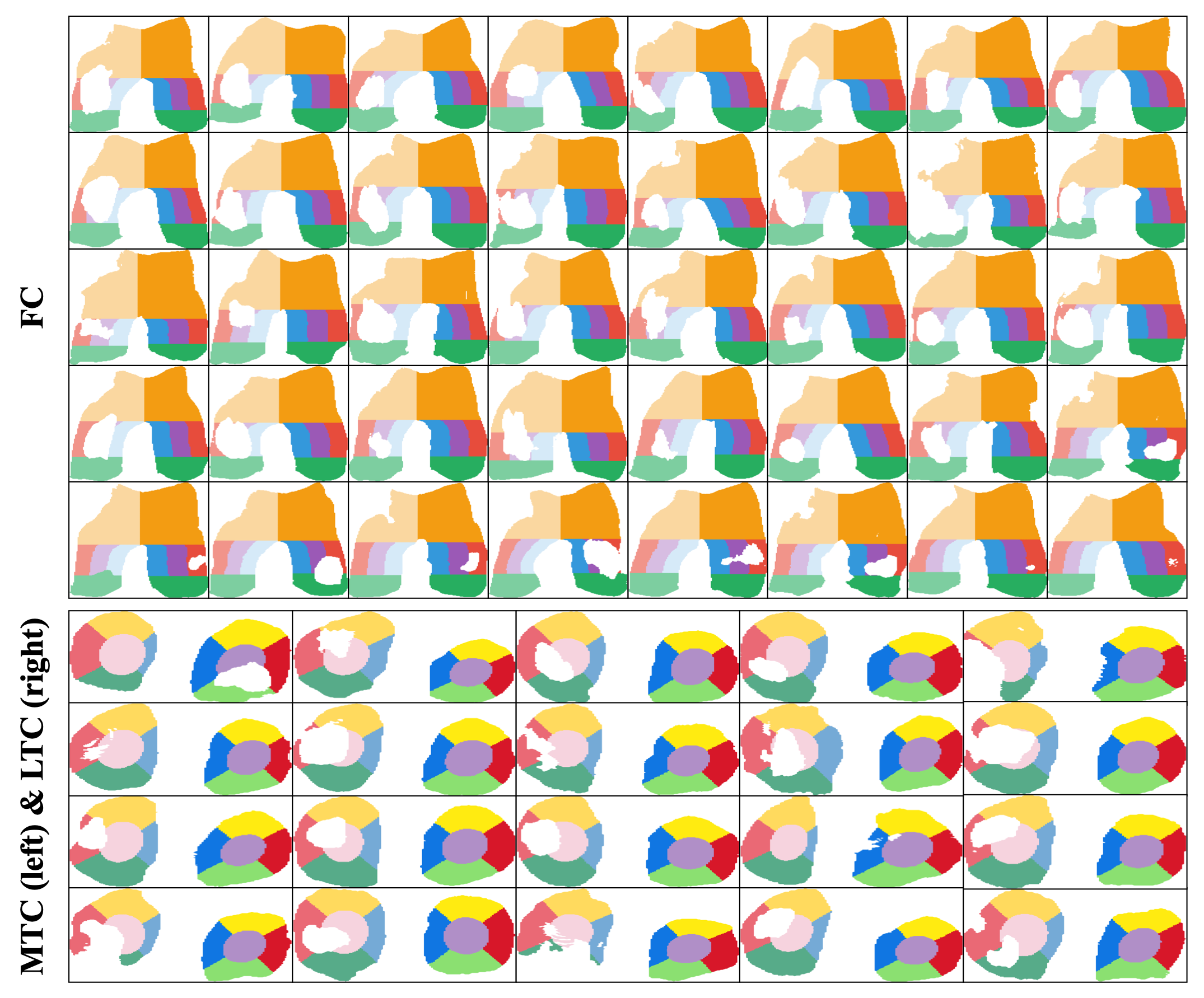}}
\caption{Cartilage subregions obtained from our rule-based parcellation method. The proposed method is applied to the model segmentation to get the subregional masks. The division is consistent across subjects and robust to cartilage defects. Full-thickness cartilage loss mainly locates in the central compartments of the medial femoral cartilage (cMFC) and the medial tibial cartilage (MTC). Only the results for subjects whose Kellgren–Lawrence grades are 4 are visualized.}
\label{fig:parcellationExample}
\end{figure}

Fig. \ref{fig:parcellationExample} shows examples of the rule-based cartilage parcellation. To emphasize that our method is robust to severe cartilage defects, we visualized the cartilage subregions of subjects whose Kellgren--Lawrence (KL) grades are 4. The cartilage division is shown to be consistent across subjects and robust to FCL. Additionally, Fig. \ref{fig:parcellationExample} suggests that FCL mainly locates in medial regions, such as cMFC and MTC.

\subsection{Regional Quantification}
\label{subsec:method-regionalQuant}
In this work, we quantified the mean thickness over the reconstructed surface,
which is equivalent to the definition of the \enquote{cartilage thickness over
total subchondral bone area} (\emph{i.e.}, ThCtAB) adopted in a previous
study\cite[Tab. II]{eckstein2006proposal}. The surface area of a cartilage
region is defined as the summation of triangular areas. The cartilage volume is
defined as the multiplicative product of the voxel volume and voxel number. The
proposed framework outputs the mean thickness, FCL in percentage, surface area,
and volume for each cartilage subregion.

\subsection{Implementation Details}
\label{subsec:method-implementation}
The segmentation, registration, and template construction models were implemented in Python (3.10) with external libraries including voxelmorph (0.2), nnunet (1.7.0), keras (2.8.0), and simpleitk (2.1.1.2). The trained models were integrated into the proposed framework implemented in Matlab (2021b, Natick, Massachusetts, U.S.A.).

\section{Experiment}
\label{sec:experiment}

\subsection{Data}
\label{subsec:experiment-data}

We used a subset of the public Osteoarthritis Initiative (OAI) dataset released
by Zuse Institute Berlin (OAI-ZIB). The OAI-ZIB dataset contains right knee MR
images and manual segmentation of the femoral cartilage, tibial cartilage,
femur, and tibia for 507 subjects. We split the manual segmentation of tibial
cartilage into MTC and LTC using an in-house algorithm. MR images were acquired
from the fat-suppressed 3D dual-echo in steady state (DESS) sequence. The
reconstructed image resolution is $[0.7, 0.36, 0.36]~{\rm mm}$. Details of the MR protocol are
given in an OAI report\cite{peterfy2008osteoarthritis}. Clinical information
of subjects, such as the KL grade, was retrieved from the OAI dataset. We
successfully retrieved the KL grades for 481 subjects and sampled them into 5
subsets for various experiments. Table \ref{tab:dataset} gives the numbers and distributions of KL grades for each subset. 

\begin{table}[!ht]
\footnotesize
\caption{Datasets used in this work. The public OAI-ZIB dataset is sampled into 5 subsets for various experiments.}
\label{tab:dataset}
\setlength{\tabcolsep}{4pt}
\renewcommand{\arraystretch}{1.3}
\centering
\begin{tabular}{llll}
    \hline
    \textbf{Dataset} & \textbf{Utility (Model)} & \textbf{Size} &\textbf{KL Grade (KL0-4)}\\
    \hline\hline
    OAI-ZIB & - & 507 & 103/58/108/139/73 \\
    \hline
    dataset 1 & template construction  ($\mathcal{G}_{\bm{\theta}_t}$) & 103 & 103/0/0/0/0 \\
    dataset 2 & model training ($\mathcal{F}_{\bm{\theta}_s}$, $\mathcal{G}_{\bm{\theta}_u}$) & 383 & 82/46/86/111/58 \\
    dataset 3 & model testing ($\mathcal{F}_{\bm{\theta}_s}$, $\mathcal{G}_{\bm{\theta}_u}$) & 98 & 21/12/22/28/15 \\
    dataset 4 & framework evaluation & 481 & 103/58/108/139/73 \\
    dataset 5 & FCL manual grading & 79 & 2/1/7/22/47 \\
    \hline
\end{tabular}
\end{table}

\emph{Data for Template Construction}. We sampled 103 subjects without radiographic OA (\emph{i.e.}, with a KL grade of 0) into the template construction dataset. The MR images in this subset were used for the training of network $\mathcal{G}_{\boldsymbol{\theta}_t}$.

\emph{Data for Model Training and Testing}. We grouped 481 subjects into a training set of 383 subjects (approximately 80\%) and a testing set of 98 subjects (approximately 20\%) via stratified sampling. The segmentation network $\mathcal{F}_{\boldsymbol{\theta}_s}$ and registration network $\mathcal{G}_{\boldsymbol{\theta}_u}$ used the same training and testing sets.

\emph{Data for Framework Evaluation}. We sampled 481 subjects with valid KL grades into the framework evaluation dataset. We evaluated the proposed rule-based cartilage parcellation method and analyzed the effectiveness of the segmentation model in cartilage morphometrics using this dataset. 

\emph{Data for FCL Manual Grading}. Semiquantitative measurements of the
cartilage thickness and FCL were retrieved from the Pivotal Osteoarthritis
Initiative Magnetic Resonance Imaging Analyses (POMA) dataset.\footnote[2]{Data
retrieved from the \enquote{kmri\_poma\_tkr\_chondrometrics} non-image dataset
at https://nda.nih.gov/oai/.} The POMA dataset contains the results
of a knee replacement prediction study\cite{eckstein2013quantitative} where
the FCL metrics are quantified by Chondrometrics (Chondrometrics GmbH, Munich,
Germany). We filtered out 79 matched subjects and graded the severity of FCL
manually. With the manual grades as ground truths, we evaluated and compared
the accuracy of FCL metrics from our method and those from Chondrometrics.

\subsection{Experiment 1: Knee Template Construction}
\label{subsec:experiment-1-template}

\begin{figure}[!t]
\centerline{\includegraphics[width=0.6\columnwidth]{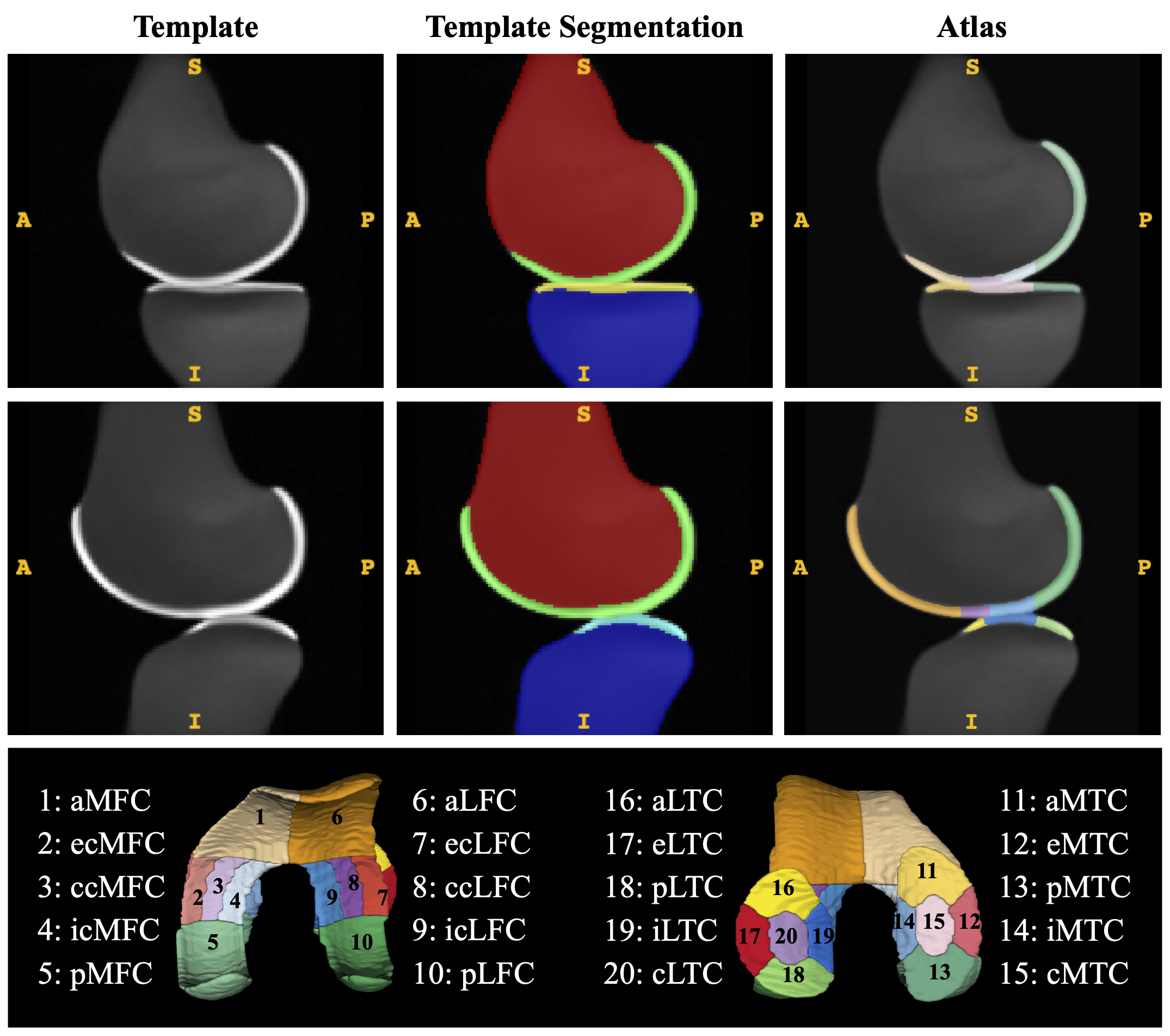}}
\caption{\enquote{CLAIR-Knee-103R}: the template, template segmentation, and atlas for the right knee. The template is an average image of bone and articular cartilage. The template segmentation consists of 5 ROIs: femur, tibia, femoral cartilage (FC), medial tibial cartilage (MTC), and lateral tibial cartilage (LTC). The atlas contains 20 subregions of cartilage. The 3D visualization of the 20-region atlas is shown in the bottom plot.}
\label{fig:template}
\end{figure}

We constructed a right-knee template, \enquote{CLAIR-Knee-103R} (Fig. \ref{fig:template}), from 103 DESS MR scans. Additionally, we created manual segmentation of the bone and cartilage for the template image, including the femur, tibia, FC, MTC, and LTC. The rule-based parcellation algorithm was applied to partition the cartilage into 20 regions, producing an atlas for the template image. The implementation details are as follows.

MR image intensities were normalized to $[0,1]$. Images in the template construction dataset $\{\boldsymbol{I}_i\}_{i=1}^{n_\text{t}}$ (Table \ref{tab:dataset}, dataset 1) were masked by respective segmentation label
in $\{\boldsymbol{S}_i^l\}_{i=1}^{n_\text{t}}$ to remove tissues other than bone and cartilage. The masked
images of size $[160,384,384]$ were downsampled with a scaling factor of 0.5 and
further cropped into a 3D volume of size $[64,128,128]$ to form a low-resolution
dataset $\{\boldsymbol{I}_i^{\text{low}}\}_{i=1}^{n_\text{t}}$. The network $\mathcal{G}_{\boldsymbol{\theta}_t}$ was trained on $\{\boldsymbol{I}_i^{\text{low}}\}_{i=1}^{n_\text{t}}$ to
learn a representative template. The hyperparameters $\{\lambda_1, \lambda_2, \lambda_3, \lambda_4\}$ in
\ref{eq:temp-3} were set at $\{0.5, 0.5, 1, 0.01\}$. The learnable template
$\hat{\boldsymbol{I}^t_i}$ in \ref{eq:temp-3} was initialized with the initial mean image:
$\hat{\boldsymbol{I}^t_0} = \bar{\boldsymbol{I}}^{\text{low}}$. The registration subnetwork $\mathcal{G}_{\boldsymbol{\theta}_v}$ in \ref{eq:temp-3}
shared the same U-Net architecture used in a previous study\cite[Fig. 3]{balakrishnan2019voxelmorph}. The network was optimized with
Adam\cite{kingma2014adam} ($lr=10^{-4}, \beta_1=0.9, \beta_2=0.999$). The model was trained for 1000
epochs with 1 volume per step and 100 steps per epoch.

\subsection{Experiment 2: Performance of Segmentation Network}
\label{subsec:experiment-2-segNet}

\begin{figure}[!t]
\centerline{\includegraphics[width=0.8\columnwidth]{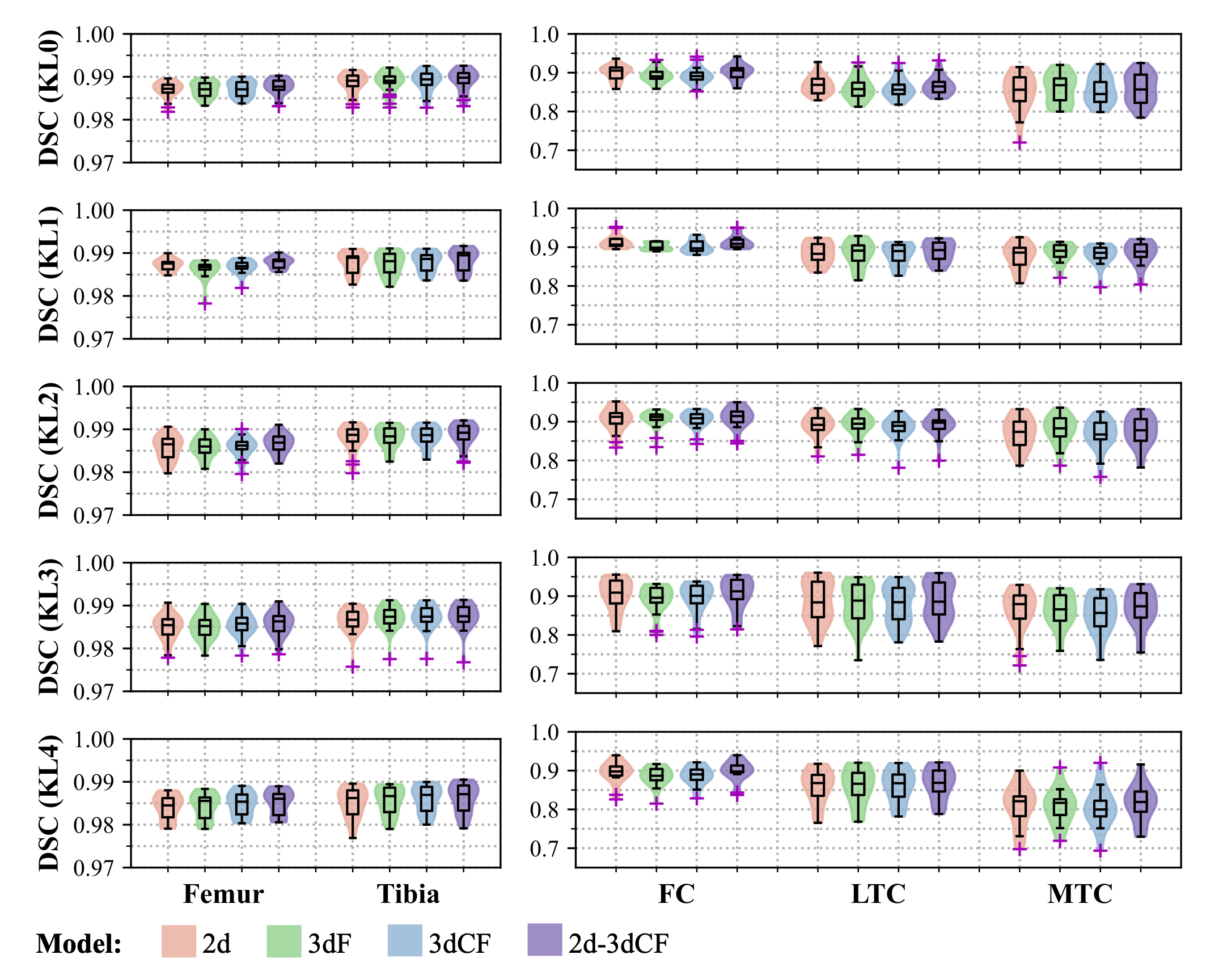}}
\caption{Model performance in tissue segmentation. The Dice similarity coefficient (DSC) is visualized for each model, KL group, and ROI.
}
\label{fig:segNet}
\end{figure}

nnU-Net variants were trained on the model training dataset (Table \ref{tab:dataset}, dataset 2) and evaluated on the model testing dataset (Table \ref{tab:dataset}, dataset 3). Network performances were evaluated by the Dice similarity coefficient (DSC). Table \ref{tab:seg} shows the performances of each model variant and the most effective model ensemble. Additionally, we visualize the network performances for each KL group and ROI in Fig. \ref{fig:segNet}. The results show that cartilage segmentation was more challenging than bone segmentation. There was a performance drop in the KL4 group for all models, especially for the MTC. This is probably because the MTC suffers from more cartilage lesions in general. All models had comparative performances, and we integrated the best model (\emph{i.e.}, 2d-3dCF) into the proposed framework.

\begin{table}[!ht]
\footnotesize
\caption{Model performances in bone and cartilage segmentation evaluated by Dice similarity coefficient (DSC). Metrics are presented as \enquote{mean value (standard deviation)}. Three nnU-Net variants and a model ensemble are evaluated. The best performances for each ROI are in bold. (2d: 2D U-Net; 3dF: full-resolution 3D U-Net; 3dCF: 3D U-Net cascade; 2d-3dCF: the ensemble of 2d and 3dCF)}
\label{tab:seg}
\setlength{\tabcolsep}{4pt}
\renewcommand{\arraystretch}{1.3}
\centering
\begin{tabular}{ccccc}
    \hline
    \textbf{ROI} & \textbf{\makecell{nnUNet \\ (2d)}} & \textbf{\makecell{nnUNet \\ (3dF)}} & \textbf{\makecell{nnUNet \\ (3dCF)}} & \textbf{\makecell{nnUNet \\ (2d-3dCF)}} \\
    \hline\hline
    Femur & 0.986 (0.003) & 0.986 (0.003) & 0.986 (0.003) & \textbf{0.987 (0.003)} \\
    Tibia & 0.987 (0.003) & 0.988 (0.003) & 0.988 (0.003) & \textbf{0.988 (0.003)} \\
    FC & 0.903 (0.030) & 0.895 (0.026) & 0.896 (0.028) & \textbf{0.906 (0.029)} \\
    LTC & 0.878 (0.040) & 0.875 (0.042) & 0.872 (0.038) & \textbf{0.881 (0.038)} \\
    MTC & 0.856 (0.051) & 0.861 (0.045) & 0.853 (0.046) & \textbf{0.861 (0.046)} \\
    \hline
\end{tabular}
\end{table}

\subsection{Experiment 3: Performance of Registration Network}
\label{subsec:experiment-3-regNet}

We trained 3 VoxelMorph variants with low-resolution images in the model training dataset (Table \ref{tab:dataset}, dataset 2): (1) a baseline model with the MSE loss, (2) a
model with MSE loss and doubled channels, and (3) a model with the LNCC loss and
doubled channels. The hyperparameter $\lambda$ in \ref{eq:reg-2} was set
at 0.01. We chose a 27-voxel neighborhood for the LNCC calculation in
\ref{eq:reg-5}: $\mid N(p)\mid =27$. The numbers of encoder--decoder filters
(\emph{i.e.}, channels) in the VoxelMorph baseline model were
$\{[16,32,32,32],[32,32,32,32,32,16,16]\}$\cite[Fig. 3]{balakrishnan2019voxelmorph}. The model was
optimized with Adam ($lr=10^{-4}, \beta_1=0.9, \beta_2=0.999$) and trained for 1000 epochs with 1 volume
per step and 100 steps per epoch.

\begin{figure}[!t]
\centerline{\includegraphics[width=0.6\columnwidth]{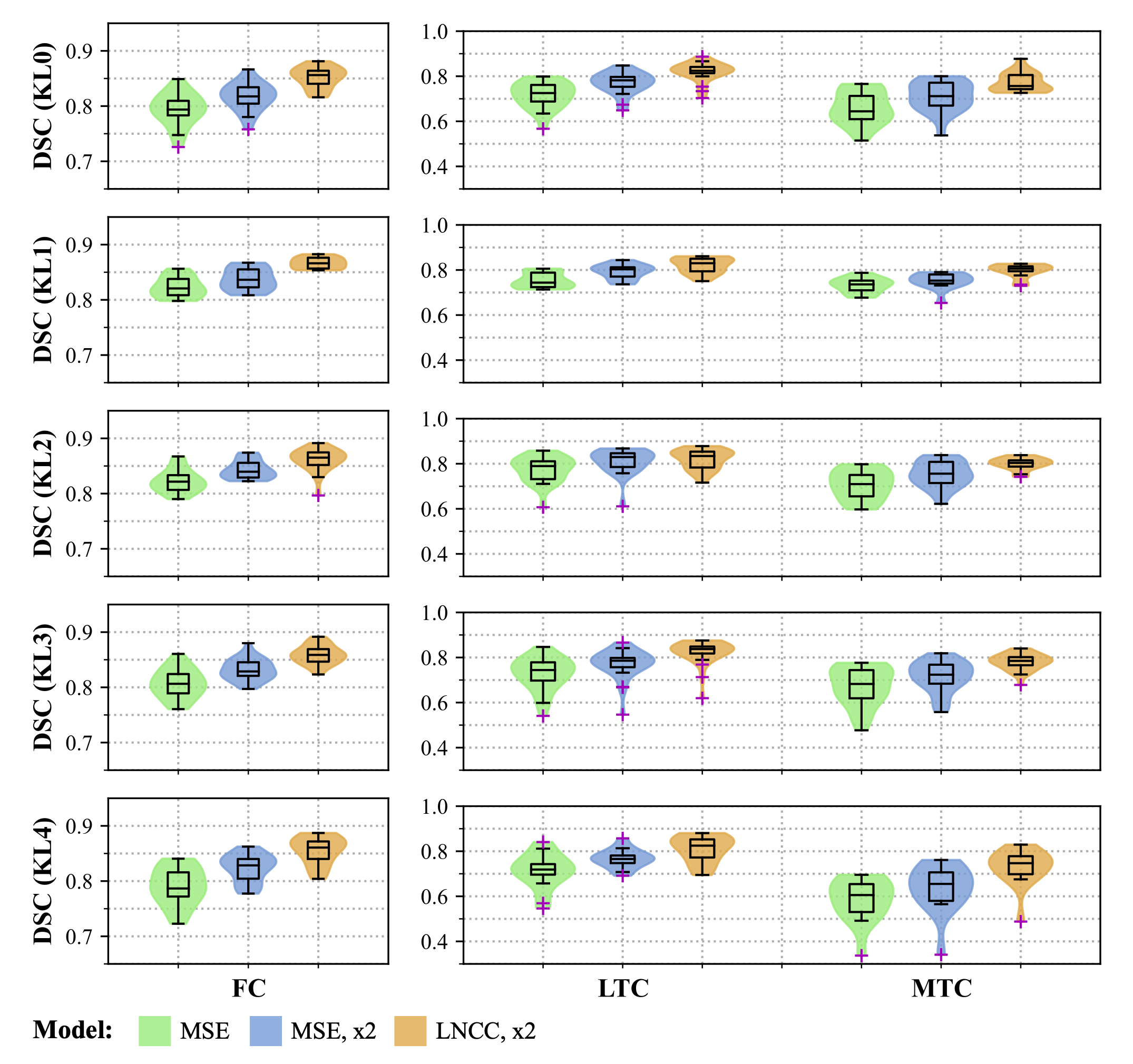}}
\caption{Model performance in template-to-image registration. The Dice similarity coefficient (DSC) is visualized for each model, KL group, and ROI.}
\label{fig:regNet}
\end{figure}

Model performances were evaluated with the low-resolution images in the model testing dataset (Table \ref{tab:dataset}, dataset 3). Model performance in image registration can be evaluated from the similarity between the moving and fixed images. Alternatively, it can be evaluated from the volume overlap between the segmentation masks of the moving and target images. In this work, we were interested in the model performance in template-to-image registration. We thus evaluated the model by calculating the DSC between the warped template segmentation mask and individual segmentation mask. Table \ref{tab:reg} gives the evaluation metric DSC for each ROI, showing that the
model with the LNCC loss and doubled channels (\emph{i.e.}, LNCC, $\times2$)
achieved the best performance. To investigate the effect of cartilage lesions
on the model performance, we visualize the DSC for each KL group and ROI in
Fig. \ref{fig:regNet}. The model with doubled channels performed better than the
baseline model as it tended to capture more fine-grained features. Consistent
with previous findings\cite{balakrishnan2019voxelmorph}, the model with the
LNCC loss was superior to the model with the MSE loss. Fig. \ref{fig:regNet} shows
that the use of LNCC loss and doubled channels improved the model performance,
especially for the MTC in the KL4 group which may suffer from more cartilage
lesions as shown in Fig. \ref{fig:parcellationExample}.

\begin{table}[!t]
\footnotesize
\caption{Model performances in template-to-image registration evaluated by Dice Similarity Coefficient (DSC). Metrics are presented as \enquote{mean value (standard deviation)}. Three VoxelMorph variants are evaluated. The best performances for each ROI are in bold. ($\times2$: doubled channels; MSE: mean squared error; LNCC: local normalized cross-correlation)}
\label{tab:reg}
\setlength{\tabcolsep}{10pt}
\renewcommand{\arraystretch}{1.3}
\centering
\begin{tabular}{cccc}
    \hline
    \textbf{ROI} & \textbf{\makecell{VoxelMorph \\ (MSE)}} & \textbf{\makecell{VoxelMorph \\ (MSE, $\bm{\times2}$)}} & \textbf{\makecell{VoxelMorph \\ (LNCC, $\bm{\times2}$)}} \\
    \hline\hline
    FC & 0.807 (0.029) & 0.831 (0.023) & \textbf{0.858 (0.019)}\\
    MTC & 0.671 (0.085) & 0.715 (0.078) & \textbf{0.777 (0.049)}\\
    LTC & 0.736 (0.067) & 0.782 (0.054) & \textbf{0.819 (0.047)}\\
    \hline
\end{tabular}
\end{table}

\subsection{Experiment 4: Rule-based \& Atlas-based Parcellation}
\label{subsec:experiment-4-parcellation}

The rule-based parcellation algorithm in the proposed framework is vital for accurate regional quantification as it accurately partitions the cartilage regardless of the variation in shape and lesions. In contrast, atlas-based parcellation methods heavily rely on accurate registration between the template and the input image. Misalignment can lead to inaccurate cartilage division. 

\begin{figure}[!t]
\centerline{\includegraphics[width=1\columnwidth]{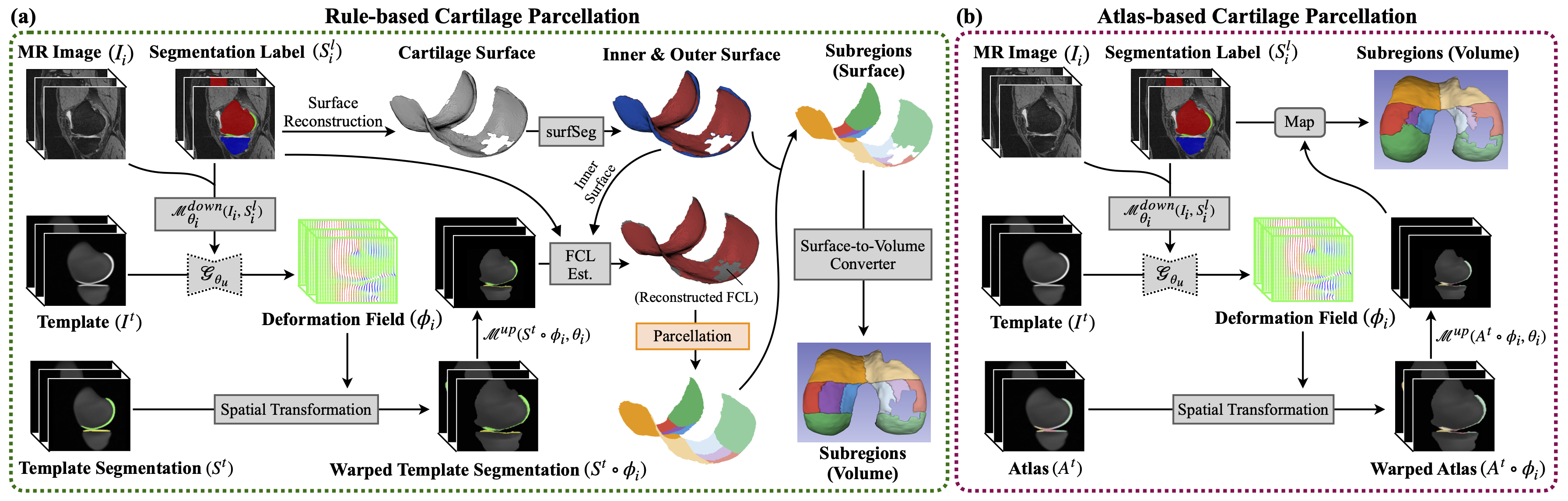}}
\caption{Image processing pipelines for comparing the rule-based and atlas-based cartilage parcellation methods. (a) The input image goes through the proposed pipeline to get a surface parcellation that is subsequently converted to a volume parcellation. (b) The atlas is aligned with the segmentation label by applying the deformation field estimated from the template-to-image registration. The subregional labels in the atlas are then transferred to a volumetric parcellation.}
\label{fig:parcellation-pipeline}
\end{figure}

In this experiment, we compared the proposed rule-based cartilage parcellation with an atlas-based method using data from the framework evaluation dataset (Table \ref{tab:dataset}, dataset 4). We took the segmentation labels $\boldsymbol{S}^l_i$ as inputs to avoid errors arising from tissue segmentation. Fig. \ref{fig:parcellation-pipeline}a and Fig. \ref{fig:parcellation-pipeline}b illustrate the pipelines of the compared methods. For the rule-based method, the input image $\boldsymbol{I}_i$ and the segmentation label $\boldsymbol{S}^l_i$ were used to generate a surface parcellation. A nearest-neighbor mapping function then converts the surface parcellation to volume parcellation. For the atlas-based method, the atlas $\boldsymbol{A}^t$ is aligned with the segmentation label $\boldsymbol{S}_i^l$ by applying the deformation field $\boldsymbol{\phi}_i$ estimated from template-to-image registration. The nearest-neighbor mapping function then transferred the subregional labels in the warped atlas $\boldsymbol{A}^t \circ \boldsymbol{\phi}_i$ to the cartilage volume. Finally, the volume overlap of subregions obtained from the two approaches was quantified by DSC. 

\begin{figure}[!t]
\centerline{\includegraphics[width=0.8\columnwidth]{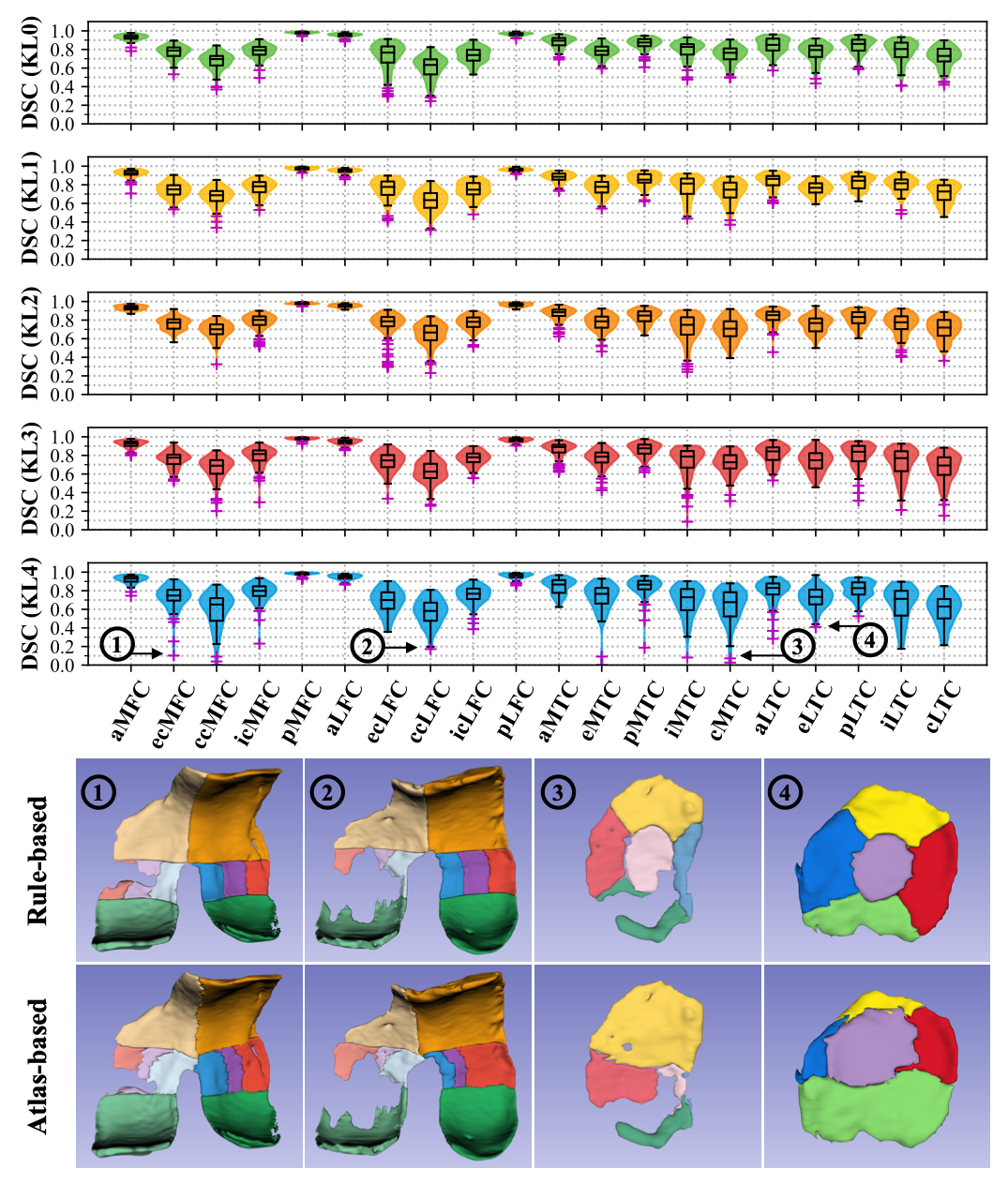}}
\caption{Volume overlap between the rule-based and atlas-based parcellation results evaluated by Dice similarity coefficient (DSC). Example results of great discrepancy between the compared methods are visualized in the bottom plots.}
\label{fig:parcellation}
\end{figure}

Fig. \ref{fig:parcellation} shows that the discrepancy between the compared methods increased as the cartilage defects increased (from KL0 to KL4). Most prominently, there is great disagreement in the KL4 group. We highlighted some of the outliers in the violin plots and visualized the volumetric parcellation obtained from each method. Visual inspection reveals that the cartilage parcellation from the proposed method is consistent with the subregion definition as shown in Fig. \ref{fig:parcellationExample} and Fig. \ref{fig:parcellation}. The atlas-based method fails to do so when there are severe cartilage lesions. The proposed rule-based method is thus superior in partitioning cartilages with severe lesions compared to the atlas-based method.

\subsection{Experiment 5: Effectiveness of Segmentation Network}
\label{subsec:experiment-5-segNetEffectiveness}

\begin{table*}[ht]
\footnotesize
\caption{Correlations of morphological quantification results from manual label and model segmentation. The Pearson's correlation coefficient $\rho$, the root mean squared deviation (RMSD), and the coefficient of variation of RMSD ($\text{CV}_\text{RMSD}$) are evaluated for each of the full-thickness cartilage loss (FCL), mean thickness, surface area, and volume.}
\label{tab:CartiMorph-accuracy}
\setlength{\tabcolsep}{2pt}
\renewcommand{\arraystretch}{1.3}
\centering
\begin{tabular}{cccccccccccccc} 
    \hline
    \multirow{2}{*}{\textbf{Label}} & \multirow{2}{*}{\textbf{ROI}} & \multicolumn{3}{c}{\textbf{FCL (\%)}} & \multicolumn{3}{c}{\textbf{Mean Thickness (mm)}} & \multicolumn{3}{c}{\textbf{Surface Area ($\textbf{mm}^\textbf{2}$)}} & \multicolumn{3}{c}{\textbf{Volume ($\textbf{mm}^\textbf{3}$)}} \\
    \cmidrule(lr){3-5} \cmidrule(lr){6-8} \cmidrule(lr){9-11} \cmidrule(lr){12-14}
    {} & {} & \multicolumn{1}{c}{\bm{$\rho$}} & \multicolumn{1}{c}{\textbf{RMSD}} & \multicolumn{1}{c}{\textbf{$\text{CV}_\text{RMSD}$}} & \multicolumn{1}{c}{\bm{$\rho$}} & \multicolumn{1}{c}{\textbf{RMSD}} & \multicolumn{1}{c}{\textbf{$\text{CV}_\text{RMSD}$}} & \multicolumn{1}{c}{\bm{$\rho$}} & \multicolumn{1}{c}{\textbf{RMSD}} & \multicolumn{1}{c}{\textbf{$\text{CV}_\text{RMSD}$}} & \multicolumn{1}{c}{\bm{$\rho$}} & \multicolumn{1}{c}{\textbf{RMSD}} & \multicolumn{1}{c}{\textbf{$\text{CV}_\text{RMSD}$}}\\
    \hline\hline
    1     & aMFC  & 0.789 & 2.457 & 0.632 & 0.886 & 0.186 & 0.068 & 0.902 & 94.297 & 0.087 & 0.948 & 220.350 & 0.088 \\
    2     & ecMFC & 0.913 & 5.523 & 0.641 & 0.906 & 0.183 & 0.107 & 0.920 & 17.929 & 0.079 & 0.944 & 35.341 & 0.112 \\
    3     & ccMFC & 0.980 & 3.823 & 0.618 & 0.961 & 0.255 & 0.101 & 0.971 & 12.545 & 0.057 & 0.962 & 61.955 & 0.126 \\
    4     & icMFC & 0.804 & 4.895 & 0.535 & 0.836 & 0.243 & 0.106 & 0.947 & 14.647 & 0.055 & 0.921 & 51.196 & 0.105 \\
    5     & pMFC  & 0.853 & 2.154 & 0.425 & 0.878 & 0.130 & 0.052 & 0.960 & 64.551 & 0.051 & 0.971 & 155.549 & 0.052 \\
    6     & aLFC  & 0.719 & 2.446 & 0.747 & 0.944 & 0.128 & 0.046 & 0.949 & 92.062 & 0.058 & 0.975 & 228.042 & 0.061 \\
    7     & ecLFC & 0.875 & 5.215 & 0.643 & 0.923 & 0.184 & 0.088 & 0.949 & 15.441 & 0.064 & 0.967 & 37.014 & 0.084 \\
    8     & ccLFC & 0.904 & 2.759 & 2.679 & 0.915 & 0.253 & 0.081 & 0.970 & 10.799 & 0.043 & 0.953 & 63.285 & 0.094 \\
    9     & icLFC & 0.644 & 3.348 & 0.712 & 0.820 & 0.188 & 0.077 & 0.958 & 12.112 & 0.046 & 0.954 & 33.853 & 0.070 \\
    10    & pLFC  & 0.861 & 2.109 & 0.455 & 0.899 & 0.128 & 0.048 & 0.965 & 57.946 & 0.051 & 0.980 & 138.507 & 0.047 \\
    11    & aMTC  & 0.693 & 4.212 & 0.684 & 0.910 & 0.195 & 0.109 & 0.865 & 27.435 & 0.081 & 0.938 & 63.621 & 0.131 \\
    12    & eMTC  & 0.817 & 7.689 & 0.829 & 0.902 & 0.174 & 0.111 & 0.895 & 20.827 & 0.092 & 0.938 & 36.985 & 0.125 \\
    13    & pMTC  & 0.825 & 5.043 & 0.676 & 0.889 & 0.146 & 0.084 & 0.837 & 30.361 & 0.090 & 0.903 & 58.824 & 0.121 \\
    14    & iMTC  & 0.458 & 7.496 & 0.753 & 0.863 & 0.243 & 0.109 & 0.838 & 22.000 & 0.116 & 0.931 & 46.686 & 0.139 \\
    15    & cMTC  & 0.923 & 3.812 & 1.657 & 0.916 & 0.244 & 0.098 & 0.835 & 27.174 & 0.105 & 0.890 & 65.274 & 0.149 \\
    16    & aLTC  & 0.710 & 4.688 & 0.805 & 0.933 & 0.170 & 0.086 & 0.832 & 28.198 & 0.103 & 0.962 & 42.894 & 0.098 \\
    17    & eLTC  & 0.585 & 6.143 & 0.901 & 0.915 & 0.150 & 0.077 & 0.895 & 19.190 & 0.083 & 0.962 & 32.746 & 0.088 \\
    18    & pLTC  & 0.786 & 3.453 & 0.857 & 0.913 & 0.175 & 0.074 & 0.836 & 28.236 & 0.100 & 0.920 & 57.588 & 0.104 \\
    19    & iLTC  & 0.423 & 6.232 & 0.732 & 0.900 & 0.290 & 0.130 & 0.873 & 21.852 & 0.097 & 0.929 & 76.452 & 0.167 \\
    20    & cLTC  & 0.922 & 3.573 & 2.664 & 0.942 & 0.326 & 0.111 & 0.824 & 26.831 & 0.112 & 0.917 & 74.654 & 0.152 \\
    \hline
\end{tabular}
\end{table*}

\begin{figure}[!t]
\centerline{\includegraphics[width=1\columnwidth]{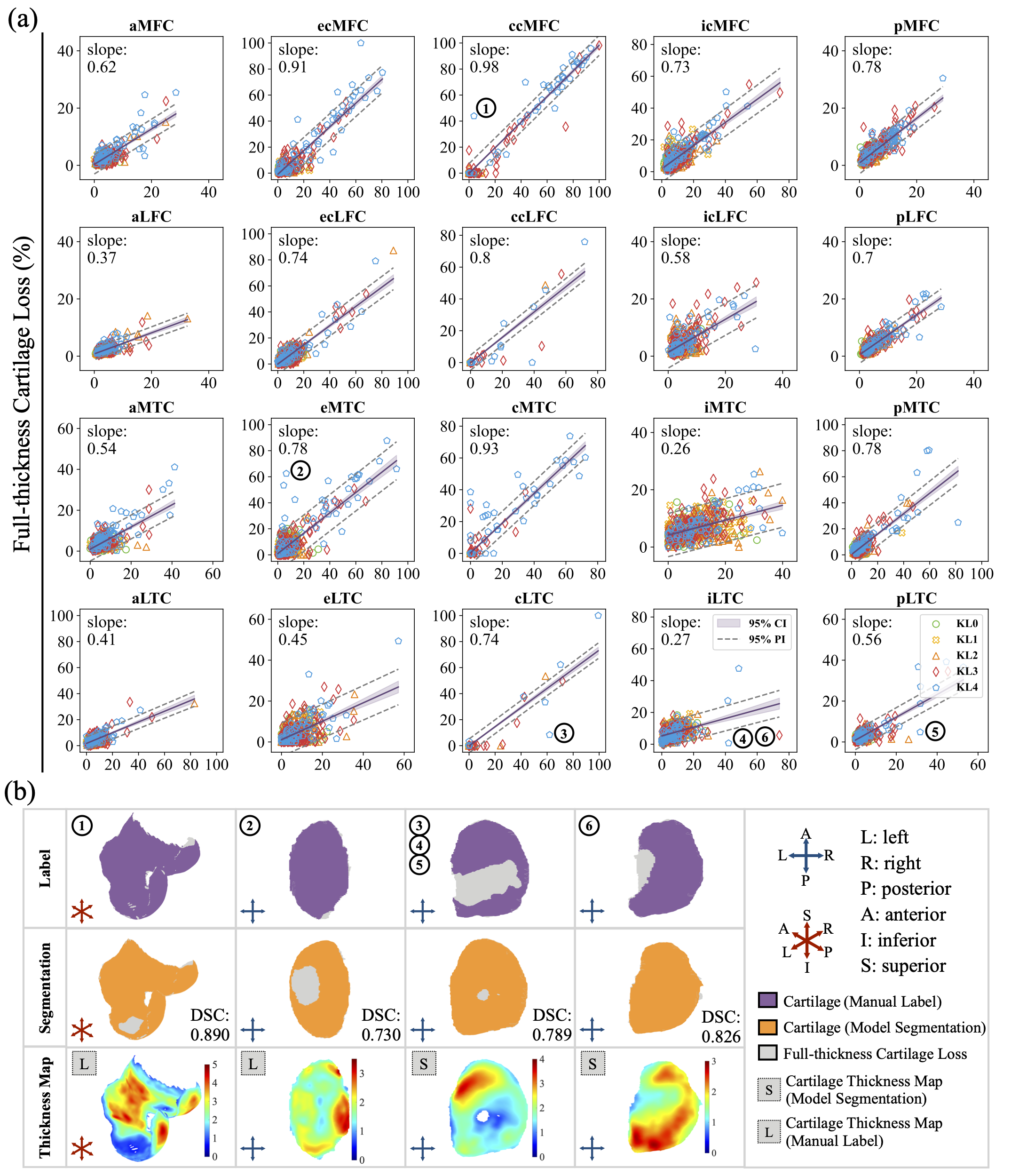}}
\caption{The effectiveness of segmentation model in full-thickness cartilage loss (FCL) estimation. (a) For each subregion, the FCL measurements gained from the manual segmentation (horizontal axis) are plotted against those from the model segmentation (vertical axis). A regression line with 95\% confident interval (CI) and 95\% prediction interval (PI) is shown in each subplot. The slope of the regression line is shown in the upper-left corner of each subplot. (b) Examples of great disagreement between the FCL measurements from manual segmentation and those from model segmentation are shown in the bottom plots in which the manual labels, model segmentation, and cartilage thickness maps are visualized.}
\label{fig:CartiMorph-accuracy-FCL}
\end{figure}

\begin{figure}[!t]
\centerline{\includegraphics[width=\columnwidth]{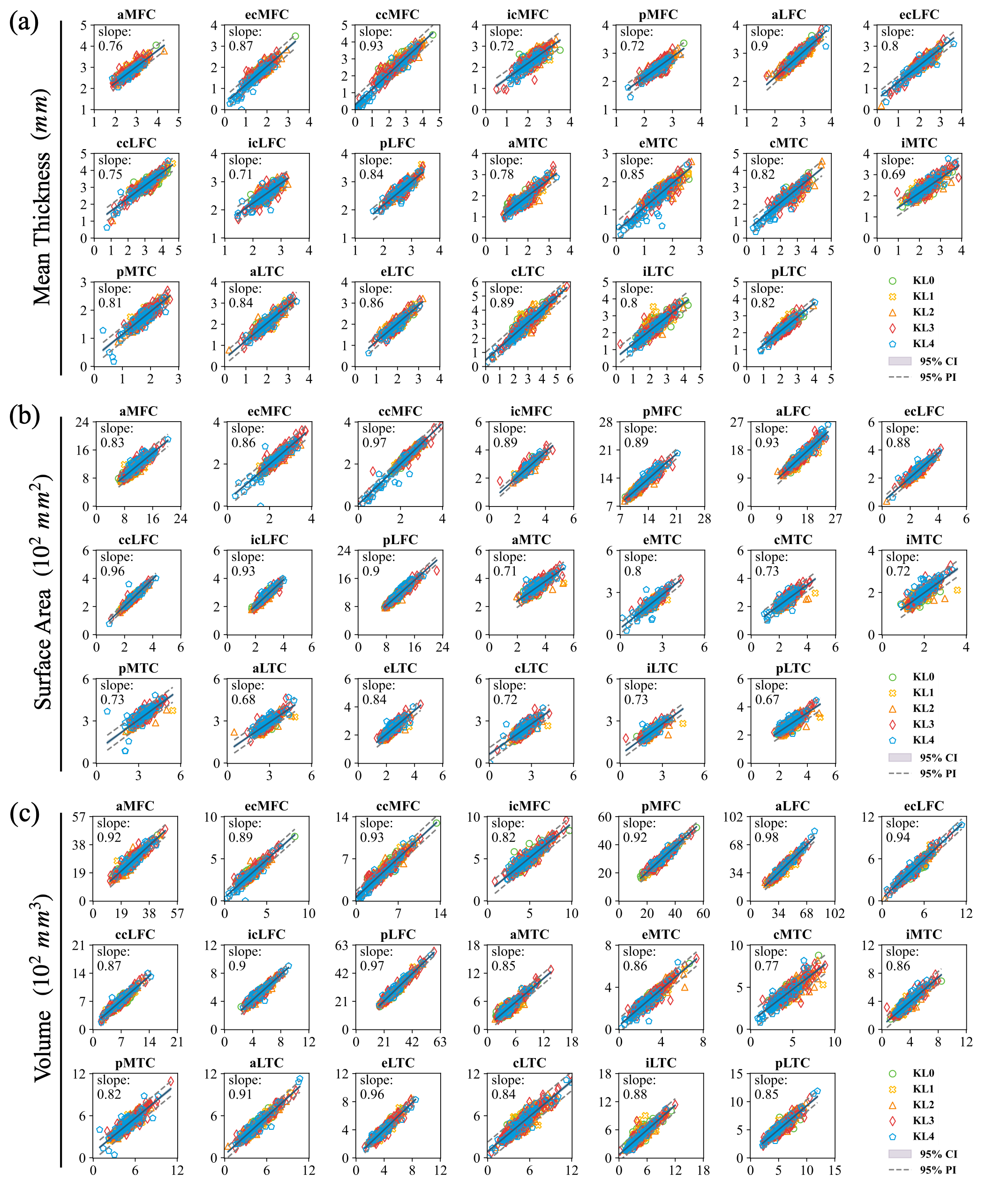}}
\caption{The effectiveness of the segmentation model in the measurements of (a) mean thickness, (b) surface area, and (c) volume. For each subregion, the results gained from the manual segmentation (horizontal axis) are plotted against those from the model segmentation (vertical axis). A regression line with 95\% confident interval (CI) and 95\% prediction interval (PI) is shown in each subplot. The slope of the regression line is shown in the upper-left corner of each subplot.}
\label{fig:CartiMorph-accuracy-3metrics}
\end{figure}

The performance of a segmentation model is usually evaluated by metrics such as
the DSC, the Jaccard similarity coefficient, precision, and recall. Pursuing
high performance in terms of such metrics is less straightforward than
evaluating the effectiveness of models in specific downstream tasks. To this
end, only a few works have compared morphological assessments made using the
outputs of deep learning models and manual
labels\cite{norman2018use,si2021knee,wirth2021accuracy,panfilov2022deep,eckstein2022detection}.
Another work\cite{tack2021towards} investigated the potential of segmentation
models in deriving quantitative features as imaging biomarkers. These works
mainly focused on the quantification of the cartilage thickness and volume. Due
to a lack of methods for automatic cartilage lesion quantification, the
effectiveness of segmentation models in lesion assessment has not been
explored. 

In this experiment, we compared quantitative measurements of the FCL, mean thickness, surface area, and volume in 20 subregions calculated from model segmentation and manual segmentation. Data from the framework evaluation dataset (Table \ref{tab:dataset}, dataset 4) were used. More specifically, the image-segmentation pairs in $\{ \{ \boldsymbol{I}_i, \boldsymbol{S}_i \}\}_{i=1}^{n_{\text{val}}}$ and the image--label pairs in $\{ \{ \boldsymbol{I}_i, \boldsymbol{S}^l_i \}\}_{i=1}^{n_{\text{val}}}$ were used for quantitative assessments. We finally analyzed the correlation of these results. 

Table \ref{tab:CartiMorph-accuracy} shows the Pearson correlation coefficient ($\rho$), the root-mean-squared deviation (RMSD), and the coefficient of variation of the RMSD ($\text{CV}_{\text{RMSD}}$). Let $\boldsymbol{Y}_s$ denotes the metrics obtained from model segmentation and $\boldsymbol{Y}_l$ denotes the metrics from manual labels. Then $\rho$, RMSD, and $\text{CV}_{\text{RMSD}}$ are calculated by
\begin{equation}
\label{eq:metric-rho}
    \rho(\boldsymbol{Y}_s, \boldsymbol{Y}_l) = \frac{cov(\boldsymbol{Y}_s, \boldsymbol{Y}_l)}{\sigma_{\boldsymbol{Y}_s} \sigma_{\boldsymbol{Y}_l}},
\end{equation}

\begin{equation}
\label{eq:metric-RMSD}
    \text{RMSD}(\boldsymbol{Y}_s, \boldsymbol{Y}_l)= \left( \frac{1}{\mid \boldsymbol{Y}_s\mid } \Vert \boldsymbol{Y}_s - \boldsymbol{Y}_l \Vert^2_2 \right) ^{1/2},
\end{equation}

\begin{equation}
\label{eq:metric-CVRMSD}
    \text{CV}_{\text{RMSD}}(\boldsymbol{Y}_s, \boldsymbol{Y}_l) = \frac{\text{RMSD}(\boldsymbol{Y}_s, \boldsymbol{Y}_l)}{\bar{\boldsymbol{Y}}_l},
\end{equation}
where $cov(\cdot,\cdot)$ denotes covariance, $\sigma$ denotes standard deviation, and $\bar{\boldsymbol{Y}}_l$ denotes the average of metrics calculated from manual labels. Despite the low correlations in some regions (\emph{e.g.}, iLTC \& iMTC), the RMSDs were less than 8\% for the FCL measurements in all subregions. Strong correlations were observed for the mean thickness ($\rho \in [0.82,0.97]$), surface area ($\rho \in [0.82,0.98]$) and volume ($\rho \in [0.89,0.98]$) measurements. 

Fig. \ref{fig:CartiMorph-accuracy-FCL} shows the FCL measurements obtained from the manual segmentation (horizontal axis) and model segmentation (vertical axis). Interestingly, there were underestimations in the FCL measurements for the anterior FC (\emph{i.e.}, aMFC and aLFC), anterior tibial cartilage (\emph{i.e.}, aMTC and aLTC), and interior tibial cartilage (\emph{i.e.}, iMTC and iLTC). To gain further insight into the discrepancies, we visualize the respective manual segmentation labels, model segmentation, and thickness maps for some outliers in the scatter plots (Fig. \ref{fig:CartiMorph-accuracy-FCL}). Outliers located at the upper-left corner of a scatter plot indicate overestimation, whereas those at the lower-right corner indicate underestimations. Fig. \ref{fig:CartiMorph-accuracy-FCL} indicates that the observed underestimations and overestimations might arise from erroneous segmentation in regions of thin cartilage. Fig. \ref{fig:CartiMorph-accuracy-3metrics} shows satisfactory reproducibility of the segmentation model in the mean thickness, surface area, and volume measurements, which indicates that these metrics are less affected by erroneous segmentation. Additionally, there are slight underestimations for these metrics (Fig. \ref{fig:CartiMorph-accuracy-3metrics}). The underestimations in these 4 metrics should be interpreted carefully and are further discussed in  \ref{subsec:discussion-effectiveness}.

\subsection{Experiment 6: Validation of FCL Estimation}
\label{subsec:experiment-6-FCLValidation}

\begin{figure}[!ht]
\centerline{\includegraphics[width=0.8\columnwidth]{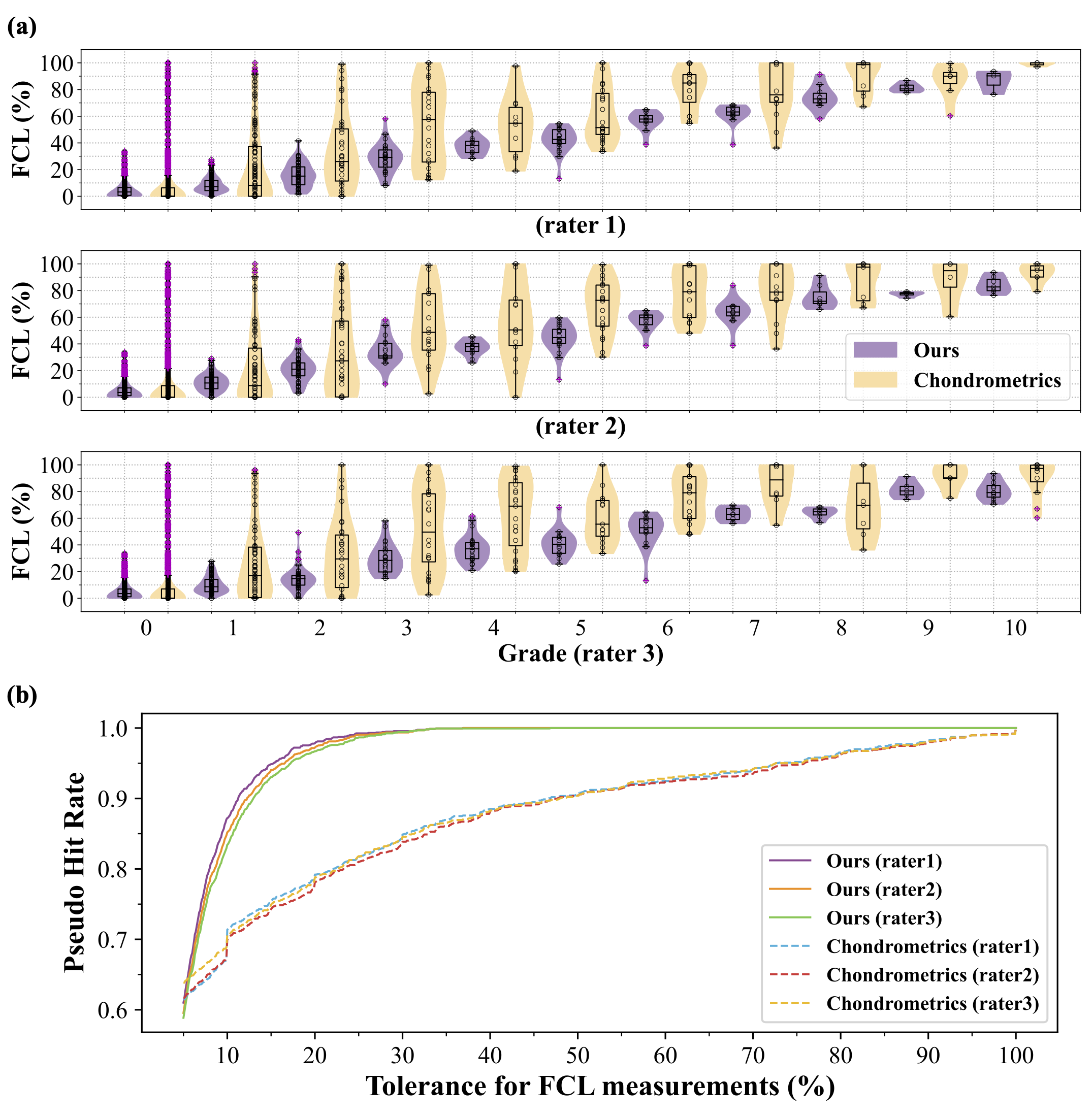}}
\caption{Comparison of the full-thickness cartilage loss (FCL) measurements from two methods. (a) The distributions of the FCL measurements in each category defined by the ground truth labels from 3 raters are visualized for each of the compared methods. (b) The pseudo hit rates (pHR) under various tolerances $\mathfrak{R}$ are visualized for the compared methods.}
\label{fig:FCL-validation}
\end{figure}

In this experiment, we compared the FCL measurements obtained from our method and those from Chondrometrics. Seventy-nine matched subjects were included (Table \ref{tab:dataset}, dataset 5). To establish a ground truth reference, 3 raters graded the FCL of each of the 20 subregions into 11 categories independently. Table \ref{tab:FLCgrade-criteria} shows the criteria for the manual grading of FCL. The manual grading results from 3 raters acted as the ground truth in 3 independent analyses. In each of them, the FCL measurements were grouped by the ground truth labels. Fig. \ref{fig:FCL-validation}a shows the distributions of the FCL measurements for each ground truth label from each rater. Compared with the Chondrometrics results, our FCL measurements had less variance and fewer outliers. The algorithm predictions of our method were thus closer to manual evaluations.

To further quantitatively evaluate the differences between the two methods, we defined the pseudo hit rate (pHR) as the percentage of hitting the ground truth within a tolerance range. Let $\tau$ be the tolerance, $\boldsymbol{Q}$ be the algorithm measurements, and $\boldsymbol{G}$ be the ground truths. The pHR is defined by
\begin{equation}
\label{eq:exp6-1}
    \text{pHR}(\boldsymbol{Q}, \boldsymbol{G} \mid  \tau) \triangleq \frac{\mid  \{q_j \mid  g_j \in [ q_j - \tau,  q_j + \tau ], \forall j \}\mid } {\mid \boldsymbol{G}\mid }.
\end{equation}
We converted the manual FCL grading labels to continuous variables by multiplying by a scaling factor of 0.1, such that grade 1 represents a ground truth value of 10\% FCL. Fig. \ref{fig:FCL-validation}b shows the pHR for the compared methods under varying tolerances of FCL in the range $[5,100]$. Note that the tolerance has the same unit as the target variable (\emph{i.e.}, the percentage of FCL in our case). We did not include tolerances less than 5\% owing to the inherited error of 5\% in the manual grading of FCL (Table \ref{tab:FLCgrade-criteria}). The Chondrometrics results have lower pHR than our method for any given tolerance level (Fig. \ref{fig:FCL-validation}b), indicating their deviations from ground truths were larger than ours (Fig. \ref{fig:FCL-validation}a). 

\begin{table}[ht]
\footnotesize
\caption{Criteria for the manual grading of full-thickness cartilage loss (FCL).}
\label{tab:FLCgrade-criteria}
\setlength{\tabcolsep}{4pt}
\renewcommand{\arraystretch}{1.3}
\centering
\begin{tabular}{cc|cc|cc|cc}
    \hline
    \textbf{Grade} & \textbf{\makecell{FCL\\(\%)}} & \textbf{Grade} & \textbf{\makecell{FCL\\(\%)}} & \textbf{Grade} & \textbf{\makecell{FCL\\(\%)}} & \textbf{Grade} & \textbf{\makecell{FCL\\(\%)}} \\
    \hline \hline
    0 & $<$5 & 3 & 30 & 6 & 60 & 9 & 90 \\
    1 & 10 & 4 & 40 & 7 & 70 & 10 & $>$95 \\
    2 & 20 & 5 & 50 & 8 & 80 &  & \\
    \hline
\end{tabular}
\end{table}

\section{Discussion}
\label{sec:discussion}

\subsection{Novelty of Rule-based Cartilage Parcellation}
\label{subsec:discussion-parcellation}

We evaluated the performance of the proposed rule-based cartilage parcellation method through visual inspection and observed that the division was consistent with the subregion definition (Fig. \ref{fig:parcellationExample}). We emphasize that the novelty of the proposed method resides in the combination of FCL estimation and the rule-based parcellation algorithm, which makes the method robust to severe cartilage loss (Fig. \ref{fig:parcellation}). Instead of simply automating an existing parcellation scheme, we designed the cartilage parcellation method for accurate regional quantification. The proposed method also benefits intrasubject and intersubject comparisons as it accurately partitions the cartilage despite the extent of cartilage lesions and the variation in cartilage shape.

\subsection{Effectiveness of Segmentation Model}
\label{subsec:discussion-effectiveness}

We quantified the effectiveness of the segmentation model in cartilage morphometrics. The metrics calculated from model segmentations were generally strongly correlated with those from manual segmentation (Table \ref{tab:CartiMorph-accuracy}). The RMSDs of the FCL measurements were within 8\%. The RMSDs of the regional mean thickness measurements were less than the voxel size of 0.36~mm. The $\text{CV}_{\text{RMSD}}$ values were less than 0.13 for the mean thickness measurements, less than 0.12 for the surface area measurements, and less than 0.17 for the volume measurements. However, there were underestimations of these regional metrics. Fig. \ref{fig:CartiMorph-accuracy-FCL} shows that erroneous segmentation in regions with thin cartilage may result in overestimations or underestimations in FCL measurements. In contrast, the mean thickness, surface area, and volume measurements were less affected by such errors (Fig. \ref{fig:CartiMorph-accuracy-3metrics}). By comparing the slopes of the regression lines in Fig. \ref{fig:CartiMorph-accuracy-FCL} and Fig. \ref{fig:CartiMorph-accuracy-3metrics}, it seems that the underestimations in FCL measurements were more severe than those in the other metrics. Yet, the slopes should be interpreted cautiously because of the differences in data distributions. Since the mean thickness, surface area, and volume are natural properties of cartilage, their measurements are close to the normal distributions, and the regression lines are less affected by outliers. Differently, the FCL is a quantitative metric of cartilage lesions. Most regional FCL measurements are near 0, and only the subjects in the KL3 and KL4 groups show greater extents of FCL. As shown in Fig. \ref{fig:CartiMorph-accuracy-FCL}, the distributions of the FCL measurements are skewed toward 0, with only a few data scattered in the positive tail of the distribution curves. Consequently, outliers might reduce the validity of the regression lines in Fig. \ref{fig:CartiMorph-accuracy-FCL}.

Our results provide new insights into the effectiveness of deep-learning-based segmentation models in cartilage morphometrics.  Deep learning models classify voxels of high certainty into the foreground class. Voxels on the boundaries usually have high uncertainty and might be misclassified as the background, which can result in underestimations in automatic cartilage morphometrics. A larger and more diverse dataset that includes subjects with severe cartilage lesions should be used to validate FCL estimation methods. As volume-based metrics such as DSC are less sensitive to the misclassification of thin regions, a model that enforces topological constraints may benefit automatic FCL estimation.

\subsection{Pseudo Hit Rate}
\label{subsec:discussion-pHR}

We evaluated the accuracy of FCL measurements using the proposed pHR and compared our metrics with those from Chondrometrics. Fig. \ref{fig:FCL-validation} shows a pHR of approximately 0.85 under the tolerance of 10\% for our method, indicating that 85\% of our FCL measurements are within the 10\% tolerance range of ground truths. Let $q_j$ be an FCL measurement made using our algorithm, $g_j$ be the ground truth, and $\tau_{0.1}$ be the tolerance level set to 10\%. It can also be interpreted as the probability that the range $[q_j - \tau_{0.1}, q_j + \tau_{0.1}]$ covers the ground truth $g_j$ is 85\% for our algorithm. Likewise, we can be 95\% confident that the range $[q_j - \tau_{0.15}, q_j + \tau_{0.15}]$ encompasses the ground truth.

The pHR is a metric that quantifies the accuracy of algorithm predictions given an acceptable tolerance. Compared with error measurements such as the RMSD, the proposed pHR is more robust to extreme outliers. More importantly, the pHR can quantify the prediction accuracy under specified tolerance. It can therefore serve as a complementary evaluation metric of algorithm performance.

\subsection{Limitations \& Future Work}
\label{subsec:discussion-limitation}

Deep learning models with complex network structures or training strategies for tissue segmentation \cite{tan2019collaborative,xu2019deepatlas,gaj2020automated,khan2022deep}, template construction \cite{dey2021generative,he2021learning,chen2021construction,pei2021learning,sinclair2022atlas}, and registration \cite{zhang2018inverse,dalca2019unsupervised,krebs2019learning,shen2019networks,mok2020fast,kim2021cyclemorph,ding2022aladdin,chen2022transmorph,shi2022xmorpher,zhu2022swin} have recently been proposed. In the present work, we did not evaluate the performances of a wide variety of deep-learning models. It is flexible to change and integrate these models into the proposed framework. We focused on establishing an automated image analysis framework that can facilitate clinical practice. Specifically, the proposed framework has the potential to assist in imaging biomarkers discovery\cite{everhart2019full,hunter2022multivariable}, longitudinal analyses of OA\cite{deveza2019trajectories,iriondo2021towards}, OA progression prediction\cite{turmezei2020quantitative,kwoh2020predicting},
efficacy analysis of interventional treatment\cite{jansen2022knee}, and
disease-modifying drug discovery\cite{hochberg2019effect}.

One of the limitations of this work is the use of a limited and slightly imbalanced framework evaluation dataset of 481 subjects (Table \ref{tab:dataset}, dataset 4). Only 73 subjects with a KL grade of 4 were included in our experiments (\ref{subsec:experiment-4-parcellation} to \ref{subsec:experiment-6-FCLValidation}). Since cartilage defects are more likely to appear in severe OA (\emph{i.e.}, in the KL3 and KL4 group), the regression lines in Fig. \ref{fig:CartiMorph-accuracy-FCL} may not accurately reflect the performance of the proposed method. Another limitation is that the patellar cartilage was not included in this study because of the lack of manual segmentation for patellar cartilage and patella in the OAI-ZIB dataset.

Future works that use larger datasets and high-quality labels of all articular cartilages to evaluate methods for imaging biomarkers extraction are of interest.

\section{Conclusion}
\label{sec:conclusion}
We introduced CartiMorph, an automated knee articular cartilage morphometrics framework. It encapsulates deep learning models for tissue segmentation, template construction, and image registration. Integrating deep learning models helps avoid manual segmentation and dramatically shortens the image registration time. We established methods for surface-normal-based cartilage thickness mapping, FCL estimation, and rule-based cartilage parcellation. The thickness mapping module comprises specialized algorithms for surface segmentation and thickness measurement, including surface closing, restricted surface dilation, surface normal estimation, orientation correction, and spatial smoothing. These functional elements constitute a robust and accurate cartilage thickness mapping pipeline. CartiMorph provides quantitative measurements of cartilage lesions (\emph{i.e.}, FCL) instead of predicting semiquantitative grades. The proposed FCL estimation method leverages the strong learning ability of deformable registration models for constructing a representative knee template. The template image, the 5-class tissue segmentation, and the 20-region atlas can provide prior knowledge of the anatomy of cartilage and bone. CartiMorph shows superior performance in FCL estimation and cartilage parcellation. As a framework for automated cartilage morphometrics, it may benefit knee-OA-related research via imaging biomarkers extraction.

\section{Conflict of Interest}
\label{sec:conflict}
Yongcheng Yao has patent ``System and method for articular cartilage thickness mapping and lesion quantification'' pending to Chinese University of Hong Kong. Weitian Chen has patent ``System and method for articular cartilage thickness mapping and lesion quantification'' pending to Chinese University of Hong Kong..

\section{Author Contribution}
\label{sec:author-contribution}
\textbf{Yongcheng Yao:} Conceptualization, Methodology, Software, Validation, Formal analysis, Investigation, Data Curation, Writing (Original Draft), Writing (Review \& Editing), Visualization, Supervision, Project administration. \textbf{Junru Zhong:} Investigation, Data Curation, Writing (Review \& Editing). \textbf{Liping Zhang:} Investigation, Data Curation, Writing (Review \& Editing). \textbf{Sheheryar Khan:} Conceptualization, Writing (Review \& Editing). \textbf{Weitian Chen:} Conceptualization, Resources, Writing (Review \& Editing), Supervision, Project administration, Funding acquisition.

\section{Acknowledgments}
\label{sec:acknowledgments}
This work was supported by a grant from the Innovation and Technology Commission of the Hong Kong SAR [MRP/001/18X] and a grant from the Faculty Innovation Award of the Chinese University of Hong Kong.


\bibliographystyle{unsrt}  
\bibliography{article.bib}

\end{document}